\documentclass[conference]{IEEEtran}

\def\review{0} 

\usepackage{tikz}  
\usetikzlibrary{spy}
\usepackage{color, colortbl}
\usepackage{pgfplots}
\pgfplotsset{compat=newest}
\usepackage{multicol}
\usepackage{circuitikz}
\usepackage[nolist]{acronym}
\usepgfplotslibrary{groupplots}
\usetikzlibrary{shapes,math}
\usepackage{siunitx}  
\usepackage{graphicx}
\usepgfplotslibrary{patchplots}
\usepackage{amsmath,amssymb}
\usepackage{subfig}
 \usepackage{float}
\usepackage{cite}
\usepackage{amsmath}
\usetikzlibrary{angles, quotes,calc,patterns}
\usepgfplotslibrary{fillbetween}

\usepackage{ifthen}
\usepackage{algorithm}
\usepackage[noend]{algpseudocode}
\usetikzlibrary{matrix,calc}
\usepackage{trfsigns}
\usepackage[font=small]{caption}

\usepackage{textcomp}
\usepackage{nicefrac}
\usepackage{multirow}
\usepackage{booktabs}
\usepackage[shortcuts,acronym]{glossaries}
\usepackage{bm}
\usepackage{lipsum}
\usepackage{soul}

\if\review1
\usepackage{todonotes}
\presetkeys%
    {todonotes}%
    {inline}{}
\else
\usepackage[disable]{todonotes}
\fi

\newcommand{\matr}[1]{{\mathbf{#1}}}
\newcommand{\vect}[1]{{\mathbf{#1}}}

\DeclareMathOperator*{\argmax}{arg\,max}
\DeclareMathOperator*{\argmin}{arg\,min}
\DeclareMathOperator{\diag}{diag}

\definecolor{uni_apfelgruen}{cmyk}{.5, 0, 1, 0}
\definecolor{uni_mittelblau}{cmyk}{1, 0.4, 0, 0}
\definecolor{uni_bulletblau}{RGB}{49,99,183}
\definecolor{uni_gelb}{cmyk}{0, 0.1, 1, 0}
\definecolor{uni_rot}{cmyk}{0, 1, 1, 0}
\definecolor{uniblauHell}{RGB}{0,190,255}
\definecolor{uniblauDunkel}{RGB}{0,65,145}
\definecolor{unigrau}{RGB}{51,51,51}
\definecolor{darkgray176}{RGB}{176,176,176}
\definecolor{lavenderplot}{RGB}{191,148,228}
\definecolor{coralplot}{RGB}{255,127,80}
\definecolor{cyanplot}{RGB}{37,219,168}

\newacronym{adc}{ADC}{analog-to-digital converter}
\newacronym{aoa}{AoA}{angle-of-arrival}
\newacronym{aod}{AoD}{angle-of-departure}
\newacronym{awgn}{AWGN}{additive white Gaussian noise}
\newacronym{cdf}{CDF}{cumulative distribution function}
\newacronym{cfar}{CFAR}{constant false alarm rate}
\newacronym{cir}{CIR}{channel impulse response}
\newacronym{ctf}{CTF}{channel transfer function}
\newacronym{dft}{DFT}{discrete Fourier transform}
\newacronym{dl}{DL}{deep learning}
\newacronym{isac}{ISAC}{integrated sensing and communication}
\newacronym{kf}{KF}{Kalman filter}
\newacronym{lmf}{LMF}{location management function}
\newacronym{mimo}{MIMO}{multiple-input and multiple-output}
\newacronym{ml}{ML}{maximum likelihood}
\newacronym{naf}{NAF}{normalized angular frequency}
\newacronym{nf}{NF}{noise figure}
\newacronym{ofdm}{OFDM}{orthogonal frequency-division multiplexing}
\newacronym{rcs}{RCS}{radar cross-section}
\newacronym{rmse}{RMSE}{root mean square error}
\newacronym{rx}{Rx}{receiver}
\newacronym{sap}{SAP}{sensing access point}
\newacronym{semf}{SeMF}{sensing management function}
\newacronym{snr}{SNR}{signal-to-noise ratio}
\newacronym{tx}{Tx}{transmitter}
\newacronym{ula}{ULA}{uniform linear array}

\title{Bistatic Information Fusion for Positioning and Tracking in Integrated Sensing and Communication}

\author{%
\IEEEauthorblockN{Maximilian Bauhofer\IEEEauthorrefmark{1}, Marcus Henninger\IEEEauthorrefmark{2}, Thorsten Wild\IEEEauthorrefmark{2}, Stephan ten Brink\IEEEauthorrefmark{1}, and Silvio Mandelli\IEEEauthorrefmark{2}}
\IEEEauthorblockA{\IEEEauthorrefmark{1}University of  Stuttgart, 70659 Stuttgart, Germany
\IEEEauthorrefmark{2}Nokia Bell Labs, 70469 Stuttgart, Germany}
\IEEEauthorblockA{E-mail: bauhofer@inue.uni-stuttgart.de}}

\newcommand\blfootnote[1]{%
  \begingroup
  \renewcommand\thefootnote{}\footnote{#1}%
  \addtocounter{footnote}{-1}%
  \endgroup
}

\begin{document}

\pgfplotsset{
    colormap={jet_inue}{
        rgb=(1, 1, 1)
        rgb=(0.99804, 0.99804, 0.99905)
        rgb=(0.99609, 0.99609, 0.99813)
        rgb=(0.99413, 0.99413, 0.99725)
        rgb=(0.99217, 0.99217, 0.99639)
        rgb=(0.99022, 0.99022, 0.99557)
        rgb=(0.98826, 0.98826, 0.99477)
        rgb=(0.9863, 0.9863, 0.99401)
        rgb=(0.98434, 0.98434, 0.99327)
        rgb=(0.98239, 0.98239, 0.99257)
        rgb=(0.98043, 0.98043, 0.9919)
        rgb=(0.97847, 0.97847, 0.99125)
        rgb=(0.97652, 0.97652, 0.99064)
        rgb=(0.97456, 0.97456, 0.99006)
        rgb=(0.9726, 0.9726, 0.98951)
        rgb=(0.97065, 0.97065, 0.98899)
        rgb=(0.96869, 0.96869, 0.9885)
        rgb=(0.96673, 0.96673, 0.98804)
        rgb=(0.96477, 0.96477, 0.98762)
        rgb=(0.96282, 0.96282, 0.98722)
        rgb=(0.96086, 0.96086, 0.98685)
        rgb=(0.9589, 0.9589, 0.98652)
        rgb=(0.95695, 0.95695, 0.98621)
        rgb=(0.95499, 0.95499, 0.98593)
        rgb=(0.95303, 0.95303, 0.98569)
        rgb=(0.95108, 0.95108, 0.98548)
        rgb=(0.94912, 0.94912, 0.98529)
        rgb=(0.94716, 0.94716, 0.98514)
        rgb=(0.94521, 0.94521, 0.98502)
        rgb=(0.94325, 0.94325, 0.98493)
        rgb=(0.94129, 0.94129, 0.98486)
        rgb=(0.93933, 0.93933, 0.98483)
        rgb=(0.93738, 0.93738, 0.98483)
        rgb=(0.93542, 0.93542, 0.98486)
        rgb=(0.93346, 0.93346, 0.98493)
        rgb=(0.93151, 0.93151, 0.98502)
        rgb=(0.92955, 0.92955, 0.98514)
        rgb=(0.92759, 0.92759, 0.98529)
        rgb=(0.92564, 0.92564, 0.98548)
        rgb=(0.92368, 0.92368, 0.98569)
        rgb=(0.92172, 0.92172, 0.98593)
        rgb=(0.91977, 0.91977, 0.98621)
        rgb=(0.91781, 0.91781, 0.98652)
        rgb=(0.91585, 0.91585, 0.98685)
        rgb=(0.91389, 0.91389, 0.98722)
        rgb=(0.91194, 0.91194, 0.98762)
        rgb=(0.90998, 0.90998, 0.98804)
        rgb=(0.90802, 0.90802, 0.9885)
        rgb=(0.90607, 0.90607, 0.98899)
        rgb=(0.90411, 0.90411, 0.98951)
        rgb=(0.90215, 0.90215, 0.99006)
        rgb=(0.9002, 0.9002, 0.99064)
        rgb=(0.89824, 0.89824, 0.99125)
        rgb=(0.89628, 0.89628, 0.9919)
        rgb=(0.89432, 0.89432, 0.99257)
        rgb=(0.89237, 0.89237, 0.99327)
        rgb=(0.89041, 0.89041, 0.99401)
        rgb=(0.88845, 0.88845, 0.99477)
        rgb=(0.8865, 0.8865, 0.99557)
        rgb=(0.88454, 0.88454, 0.99639)
        rgb=(0.88258, 0.88258, 0.99725)
        rgb=(0.88063, 0.88063, 0.99813)
        rgb=(0.87867, 0.87867, 0.99905)
        rgb=(0.87671, 0.87671, 1)
        rgb=(0.87476, 0.87573, 1)
        rgb=(0.8728, 0.87479, 1)
        rgb=(0.87084, 0.87387, 1)
        rgb=(0.86888, 0.87298, 1)
        rgb=(0.86693, 0.87213, 1)
        rgb=(0.86497, 0.8713, 1)
        rgb=(0.86301, 0.87051, 1)
        rgb=(0.86106, 0.86974, 1)
        rgb=(0.8591, 0.86901, 1)
        rgb=(0.85714, 0.8683, 1)
        rgb=(0.85519, 0.86763, 1)
        rgb=(0.85323, 0.86699, 1)
        rgb=(0.85127, 0.86638, 1)
        rgb=(0.84932, 0.8658, 1)
        rgb=(0.84736, 0.86525, 1)
        rgb=(0.8454, 0.86473, 1)
        rgb=(0.84344, 0.86424, 1)
        rgb=(0.84149, 0.86378, 1)
        rgb=(0.83953, 0.86335, 1)
        rgb=(0.83757, 0.86295, 1)
        rgb=(0.83562, 0.86259, 1)
        rgb=(0.83366, 0.86225, 1)
        rgb=(0.8317, 0.86194, 1)
        rgb=(0.82975, 0.86167, 1)
        rgb=(0.82779, 0.86142, 1)
        rgb=(0.82583, 0.86121, 1)
        rgb=(0.82387, 0.86103, 1)
        rgb=(0.82192, 0.86087, 1)
        rgb=(0.81996, 0.86075, 1)
        rgb=(0.818, 0.86066, 1)
        rgb=(0.81605, 0.8606, 1)
        rgb=(0.81409, 0.86057, 1)
        rgb=(0.81213, 0.86057, 1)
        rgb=(0.81018, 0.8606, 1)
        rgb=(0.80822, 0.86066, 1)
        rgb=(0.80626, 0.86075, 1)
        rgb=(0.80431, 0.86087, 1)
        rgb=(0.80235, 0.86103, 1)
        rgb=(0.80039, 0.86121, 1)
        rgb=(0.79843, 0.86142, 1)
        rgb=(0.79648, 0.86167, 1)
        rgb=(0.79452, 0.86194, 1)
        rgb=(0.79256, 0.86225, 1)
        rgb=(0.79061, 0.86259, 1)
        rgb=(0.78865, 0.86295, 1)
        rgb=(0.78669, 0.86335, 1)
        rgb=(0.78474, 0.86378, 1)
        rgb=(0.78278, 0.86424, 1)
        rgb=(0.78082, 0.86473, 1)
        rgb=(0.77886, 0.86525, 1)
        rgb=(0.77691, 0.8658, 1)
        rgb=(0.77495, 0.86638, 1)
        rgb=(0.77299, 0.86699, 1)
        rgb=(0.77104, 0.86763, 1)
        rgb=(0.76908, 0.8683, 1)
        rgb=(0.76712, 0.86901, 1)
        rgb=(0.76517, 0.86974, 1)
        rgb=(0.76321, 0.87051, 1)
        rgb=(0.76125, 0.8713, 1)
        rgb=(0.7593, 0.87213, 1)
        rgb=(0.75734, 0.87298, 1)
        rgb=(0.75538, 0.87387, 1)
        rgb=(0.75342, 0.87479, 1)
        rgb=(0.75147, 0.87573, 1)
        rgb=(0.74951, 0.87671, 1)
        rgb=(0.74755, 0.87772, 1)
        rgb=(0.7456, 0.87876, 1)
        rgb=(0.74364, 0.87983, 1)
        rgb=(0.74168, 0.88093, 1)
        rgb=(0.73973, 0.88206, 1)
        rgb=(0.73777, 0.88323, 1)
        rgb=(0.73581, 0.88442, 1)
        rgb=(0.73386, 0.88564, 1)
        rgb=(0.7319, 0.88689, 1)
        rgb=(0.72994, 0.88818, 1)
        rgb=(0.72798, 0.88949, 1)
        rgb=(0.72603, 0.89084, 1)
        rgb=(0.72407, 0.89222, 1)
        rgb=(0.72211, 0.89362, 1)
        rgb=(0.72016, 0.89506, 1)
        rgb=(0.7182, 0.89653, 1)
        rgb=(0.71624, 0.89802, 1)
        rgb=(0.71429, 0.89955, 1)
        rgb=(0.71233, 0.90111, 1)
        rgb=(0.71037, 0.9027, 1)
        rgb=(0.70841, 0.90432, 1)
        rgb=(0.70646, 0.90597, 1)
        rgb=(0.7045, 0.90766, 1)
        rgb=(0.70254, 0.90937, 1)
        rgb=(0.70059, 0.91111, 1)
        rgb=(0.69863, 0.91289, 1)
        rgb=(0.69667, 0.91469, 1)
        rgb=(0.69472, 0.91652, 1)
        rgb=(0.69276, 0.91839, 1)
        rgb=(0.6908, 0.92028, 1)
        rgb=(0.68885, 0.92221, 1)
        rgb=(0.68689, 0.92417, 1)
        rgb=(0.68493, 0.92616, 1)
        rgb=(0.68297, 0.92817, 1)
        rgb=(0.68102, 0.93022, 1)
        rgb=(0.67906, 0.9323, 1)
        rgb=(0.6771, 0.93441, 1)
        rgb=(0.67515, 0.93655, 1)
        rgb=(0.67319, 0.93872, 1)
        rgb=(0.67123, 0.94092, 1)
        rgb=(0.66928, 0.94316, 1)
        rgb=(0.66732, 0.94542, 1)
        rgb=(0.66536, 0.94771, 1)
        rgb=(0.66341, 0.95004, 1)
        rgb=(0.66145, 0.95239, 1)
        rgb=(0.65949, 0.95478, 1)
        rgb=(0.65753, 0.95719, 1)
        rgb=(0.65558, 0.95964, 1)
        rgb=(0.65362, 0.96211, 1)
        rgb=(0.65166, 0.96462, 1)
        rgb=(0.64971, 0.96716, 1)
        rgb=(0.64775, 0.96973, 1)
        rgb=(0.64579, 0.97233, 1)
        rgb=(0.64384, 0.97496, 1)
        rgb=(0.64188, 0.97762, 1)
        rgb=(0.63992, 0.98031, 1)
        rgb=(0.63796, 0.98303, 1)
        rgb=(0.63601, 0.98578, 1)
        rgb=(0.63405, 0.98856, 1)
        rgb=(0.63209, 0.99138, 1)
        rgb=(0.63014, 0.99422, 1)
        rgb=(0.62818, 0.9971, 1)
        rgb=(0.62622, 1, 1)
        rgb=(0.6272, 1, 0.99706)
        rgb=(0.62821, 1, 0.9941)
        rgb=(0.62925, 1, 0.9911)
        rgb=(0.63032, 1, 0.98807)
        rgb=(0.63142, 1, 0.98502)
        rgb=(0.63255, 1, 0.98193)
        rgb=(0.63371, 1, 0.97881)
        rgb=(0.63491, 1, 0.97566)
        rgb=(0.63613, 1, 0.97248)
        rgb=(0.63738, 1, 0.96927)
        rgb=(0.63867, 1, 0.96603)
        rgb=(0.63998, 1, 0.96276)
        rgb=(0.64133, 1, 0.95945)
        rgb=(0.6427, 1, 0.95612)
        rgb=(0.64411, 1, 0.95276)
        rgb=(0.64555, 1, 0.94936)
        rgb=(0.64702, 1, 0.94594)
        rgb=(0.64851, 1, 0.94248)
        rgb=(0.65004, 1, 0.939)
        rgb=(0.6516, 1, 0.93548)
        rgb=(0.65319, 1, 0.93193)
        rgb=(0.65481, 1, 0.92836)
        rgb=(0.65646, 1, 0.92475)
        rgb=(0.65815, 1, 0.92111)
        rgb=(0.65986, 1, 0.91744)
        rgb=(0.6616, 1, 0.91374)
        rgb=(0.66337, 1, 0.91001)
        rgb=(0.66518, 1, 0.90625)
        rgb=(0.66701, 1, 0.90246)
        rgb=(0.66888, 1, 0.89864)
        rgb=(0.67077, 1, 0.89478)
        rgb=(0.6727, 1, 0.8909)
        rgb=(0.67466, 1, 0.88699)
        rgb=(0.67665, 1, 0.88304)
        rgb=(0.67866, 1, 0.87907)
        rgb=(0.68071, 1, 0.87506)
        rgb=(0.68279, 1, 0.87102)
        rgb=(0.6849, 1, 0.86696)
        rgb=(0.68704, 1, 0.86286)
        rgb=(0.68921, 1, 0.85873)
        rgb=(0.69141, 1, 0.85457)
        rgb=(0.69365, 1, 0.85039)
        rgb=(0.69591, 1, 0.84617)
        rgb=(0.6982, 1, 0.84192)
        rgb=(0.70053, 1, 0.83763)
        rgb=(0.70288, 1, 0.83332)
        rgb=(0.70527, 1, 0.82898)
        rgb=(0.70768, 1, 0.82461)
        rgb=(0.71013, 1, 0.82021)
        rgb=(0.7126, 1, 0.81577)
        rgb=(0.71511, 1, 0.81131)
        rgb=(0.71765, 1, 0.80681)
        rgb=(0.72022, 1, 0.80229)
        rgb=(0.72282, 1, 0.79773)
        rgb=(0.72545, 1, 0.79314)
        rgb=(0.72811, 1, 0.78853)
        rgb=(0.7308, 1, 0.78388)
        rgb=(0.73352, 1, 0.7792)
        rgb=(0.73627, 1, 0.77449)
        rgb=(0.73905, 1, 0.76975)
        rgb=(0.74187, 1, 0.76498)
        rgb=(0.74471, 1, 0.76018)
        rgb=(0.74758, 1, 0.75535)
        rgb=(0.75049, 1, 0.75049)
        rgb=(0.75342, 1, 0.7456)
        rgb=(0.75639, 1, 0.74067)
        rgb=(0.75939, 1, 0.73572)
        rgb=(0.76241, 1, 0.73074)
        rgb=(0.76547, 1, 0.72572)
        rgb=(0.76856, 1, 0.72068)
        rgb=(0.77168, 1, 0.7156)
        rgb=(0.77483, 1, 0.71049)
        rgb=(0.77801, 1, 0.70536)
        rgb=(0.78122, 1, 0.70019)
        rgb=(0.78446, 1, 0.69499)
        rgb=(0.78773, 1, 0.68976)
        rgb=(0.79103, 1, 0.6845)
        rgb=(0.79437, 1, 0.67921)
        rgb=(0.79773, 1, 0.67389)
        rgb=(0.80113, 1, 0.66854)
        rgb=(0.80455, 1, 0.66316)
        rgb=(0.80801, 1, 0.65775)
        rgb=(0.81149, 1, 0.65231)
        rgb=(0.81501, 1, 0.64683)
        rgb=(0.81855, 1, 0.64133)
        rgb=(0.82213, 1, 0.63579)
        rgb=(0.82574, 1, 0.63023)
        rgb=(0.82938, 1, 0.62463)
        rgb=(0.83305, 1, 0.61901)
        rgb=(0.83675, 1, 0.61335)
        rgb=(0.84048, 1, 0.60766)
        rgb=(0.84424, 1, 0.60194)
        rgb=(0.84803, 1, 0.5962)
        rgb=(0.85185, 1, 0.59042)
        rgb=(0.85571, 1, 0.58461)
        rgb=(0.85959, 1, 0.57877)
        rgb=(0.8635, 1, 0.5729)
        rgb=(0.86745, 1, 0.56699)
        rgb=(0.87142, 1, 0.56106)
        rgb=(0.87543, 1, 0.5551)
        rgb=(0.87946, 1, 0.54911)
        rgb=(0.88353, 1, 0.54308)
        rgb=(0.88763, 1, 0.53703)
        rgb=(0.89176, 1, 0.53094)
        rgb=(0.89591, 1, 0.52483)
        rgb=(0.9001, 1, 0.51868)
        rgb=(0.90432, 1, 0.51251)
        rgb=(0.90857, 1, 0.5063)
        rgb=(0.91285, 1, 0.50006)
        rgb=(0.91717, 1, 0.49379)
        rgb=(0.92151, 1, 0.48749)
        rgb=(0.92588, 1, 0.48116)
        rgb=(0.93028, 1, 0.4748)
        rgb=(0.93472, 1, 0.46841)
        rgb=(0.93918, 1, 0.46199)
        rgb=(0.94368, 1, 0.45554)
        rgb=(0.9482, 1, 0.44906)
        rgb=(0.95276, 1, 0.44255)
        rgb=(0.95734, 1, 0.436)
        rgb=(0.96196, 1, 0.42943)
        rgb=(0.96661, 1, 0.42282)
        rgb=(0.97129, 1, 0.41619)
        rgb=(0.976, 1, 0.40952)
        rgb=(0.98074, 1, 0.40283)
        rgb=(0.98551, 1, 0.3961)
        rgb=(0.99031, 1, 0.38934)
        rgb=(0.99514, 1, 0.38255)
        rgb=(1, 1, 0.37573)
        rgb=(1, 0.99511, 0.37378)
        rgb=(1, 0.99018, 0.37182)
        rgb=(1, 0.98523, 0.36986)
        rgb=(1, 0.98025, 0.36791)
        rgb=(1, 0.97523, 0.36595)
        rgb=(1, 0.97019, 0.36399)
        rgb=(1, 0.96511, 0.36204)
        rgb=(1, 0.96, 0.36008)
        rgb=(1, 0.95487, 0.35812)
        rgb=(1, 0.9497, 0.35616)
        rgb=(1, 0.9445, 0.35421)
        rgb=(1, 0.93927, 0.35225)
        rgb=(1, 0.93401, 0.35029)
        rgb=(1, 0.92872, 0.34834)
        rgb=(1, 0.9234, 0.34638)
        rgb=(1, 0.91805, 0.34442)
        rgb=(1, 0.91267, 0.34247)
        rgb=(1, 0.90726, 0.34051)
        rgb=(1, 0.90182, 0.33855)
        rgb=(1, 0.89634, 0.33659)
        rgb=(1, 0.89084, 0.33464)
        rgb=(1, 0.8853, 0.33268)
        rgb=(1, 0.87974, 0.33072)
        rgb=(1, 0.87414, 0.32877)
        rgb=(1, 0.86852, 0.32681)
        rgb=(1, 0.86286, 0.32485)
        rgb=(1, 0.85717, 0.3229)
        rgb=(1, 0.85146, 0.32094)
        rgb=(1, 0.84571, 0.31898)
        rgb=(1, 0.83993, 0.31703)
        rgb=(1, 0.83412, 0.31507)
        rgb=(1, 0.82828, 0.31311)
        rgb=(1, 0.82241, 0.31115)
        rgb=(1, 0.81651, 0.3092)
        rgb=(1, 0.81057, 0.30724)
        rgb=(1, 0.80461, 0.30528)
        rgb=(1, 0.79862, 0.30333)
        rgb=(1, 0.79259, 0.30137)
        rgb=(1, 0.78654, 0.29941)
        rgb=(1, 0.78045, 0.29746)
        rgb=(1, 0.77434, 0.2955)
        rgb=(1, 0.76819, 0.29354)
        rgb=(1, 0.76202, 0.29159)
        rgb=(1, 0.75581, 0.28963)
        rgb=(1, 0.74957, 0.28767)
        rgb=(1, 0.7433, 0.28571)
        rgb=(1, 0.737, 0.28376)
        rgb=(1, 0.73068, 0.2818)
        rgb=(1, 0.72432, 0.27984)
        rgb=(1, 0.71792, 0.27789)
        rgb=(1, 0.7115, 0.27593)
        rgb=(1, 0.70505, 0.27397)
        rgb=(1, 0.69857, 0.27202)
        rgb=(1, 0.69206, 0.27006)
        rgb=(1, 0.68551, 0.2681)
        rgb=(1, 0.67894, 0.26614)
        rgb=(1, 0.67233, 0.26419)
        rgb=(1, 0.6657, 0.26223)
        rgb=(1, 0.65903, 0.26027)
        rgb=(1, 0.65234, 0.25832)
        rgb=(1, 0.64561, 0.25636)
        rgb=(1, 0.63885, 0.2544)
        rgb=(1, 0.63206, 0.25245)
        rgb=(1, 0.62524, 0.25049)
        rgb=(1, 0.6184, 0.24853)
        rgb=(1, 0.61152, 0.24658)
        rgb=(1, 0.6046, 0.24462)
        rgb=(1, 0.59766, 0.24266)
        rgb=(1, 0.59069, 0.2407)
        rgb=(1, 0.58369, 0.23875)
        rgb=(1, 0.57666, 0.23679)
        rgb=(1, 0.56959, 0.23483)
        rgb=(1, 0.5625, 0.23288)
        rgb=(1, 0.55538, 0.23092)
        rgb=(1, 0.54822, 0.22896)
        rgb=(1, 0.54103, 0.22701)
        rgb=(1, 0.53382, 0.22505)
        rgb=(1, 0.52657, 0.22309)
        rgb=(1, 0.51929, 0.22114)
        rgb=(1, 0.51199, 0.21918)
        rgb=(1, 0.50465, 0.21722)
        rgb=(1, 0.49728, 0.21526)
        rgb=(1, 0.48988, 0.21331)
        rgb=(1, 0.48245, 0.21135)
        rgb=(1, 0.47499, 0.20939)
        rgb=(1, 0.4675, 0.20744)
        rgb=(1, 0.45997, 0.20548)
        rgb=(1, 0.45242, 0.20352)
        rgb=(1, 0.44484, 0.20157)
        rgb=(1, 0.43722, 0.19961)
        rgb=(1, 0.42958, 0.19765)
        rgb=(1, 0.42191, 0.19569)
        rgb=(1, 0.4142, 0.19374)
        rgb=(1, 0.40646, 0.19178)
        rgb=(1, 0.3987, 0.18982)
        rgb=(1, 0.3909, 0.18787)
        rgb=(1, 0.38307, 0.18591)
        rgb=(1, 0.37521, 0.18395)
        rgb=(1, 0.36733, 0.182)
        rgb=(1, 0.35941, 0.18004)
        rgb=(1, 0.35146, 0.17808)
        rgb=(1, 0.34347, 0.17613)
        rgb=(1, 0.33546, 0.17417)
        rgb=(1, 0.32742, 0.17221)
        rgb=(1, 0.31935, 0.17025)
        rgb=(1, 0.31125, 0.1683)
        rgb=(1, 0.30311, 0.16634)
        rgb=(1, 0.29495, 0.16438)
        rgb=(1, 0.28675, 0.16243)
        rgb=(1, 0.27853, 0.16047)
        rgb=(1, 0.27027, 0.15851)
        rgb=(1, 0.26199, 0.15656)
        rgb=(1, 0.25367, 0.1546)
        rgb=(1, 0.24532, 0.15264)
        rgb=(1, 0.23694, 0.15068)
        rgb=(1, 0.22853, 0.14873)
        rgb=(1, 0.2201, 0.14677)
        rgb=(1, 0.21163, 0.14481)
        rgb=(1, 0.20312, 0.14286)
        rgb=(1, 0.19459, 0.1409)
        rgb=(1, 0.18603, 0.13894)
        rgb=(1, 0.17744, 0.13699)
        rgb=(1, 0.16882, 0.13503)
        rgb=(1, 0.16016, 0.13307)
        rgb=(1, 0.15148, 0.13112)
        rgb=(1, 0.14277, 0.12916)
        rgb=(1, 0.13402, 0.1272)
        rgb=(1, 0.12524, 0.12524)
        rgb=(0.99315, 0.12329, 0.12329)
        rgb=(0.98627, 0.12133, 0.12133)
        rgb=(0.97936, 0.11937, 0.11937)
        rgb=(0.97242, 0.11742, 0.11742)
        rgb=(0.96545, 0.11546, 0.11546)
        rgb=(0.95845, 0.1135, 0.1135)
        rgb=(0.95141, 0.11155, 0.11155)
        rgb=(0.94435, 0.10959, 0.10959)
        rgb=(0.93726, 0.10763, 0.10763)
        rgb=(0.93013, 0.10568, 0.10568)
        rgb=(0.92298, 0.10372, 0.10372)
        rgb=(0.91579, 0.10176, 0.10176)
        rgb=(0.90857, 0.099804, 0.099804)
        rgb=(0.90133, 0.097847, 0.097847)
        rgb=(0.89405, 0.09589, 0.09589)
        rgb=(0.88674, 0.093933, 0.093933)
        rgb=(0.8794, 0.091977, 0.091977)
        rgb=(0.87203, 0.09002, 0.09002)
        rgb=(0.86463, 0.088063, 0.088063)
        rgb=(0.8572, 0.086106, 0.086106)
        rgb=(0.84974, 0.084149, 0.084149)
        rgb=(0.84225, 0.082192, 0.082192)
        rgb=(0.83473, 0.080235, 0.080235)
        rgb=(0.82718, 0.078278, 0.078278)
        rgb=(0.81959, 0.076321, 0.076321)
        rgb=(0.81198, 0.074364, 0.074364)
        rgb=(0.80434, 0.072407, 0.072407)
        rgb=(0.79666, 0.07045, 0.07045)
        rgb=(0.78896, 0.068493, 0.068493)
        rgb=(0.78122, 0.066536, 0.066536)
        rgb=(0.77345, 0.064579, 0.064579)
        rgb=(0.76566, 0.062622, 0.062622)
        rgb=(0.75783, 0.060665, 0.060665)
        rgb=(0.74997, 0.058708, 0.058708)
        rgb=(0.74208, 0.056751, 0.056751)
        rgb=(0.73416, 0.054795, 0.054795)
        rgb=(0.72621, 0.052838, 0.052838)
        rgb=(0.71823, 0.050881, 0.050881)
        rgb=(0.71022, 0.048924, 0.048924)
        rgb=(0.70218, 0.046967, 0.046967)
        rgb=(0.6941, 0.04501, 0.04501)
        rgb=(0.686, 0.043053, 0.043053)
        rgb=(0.67787, 0.041096, 0.041096)
        rgb=(0.6697, 0.039139, 0.039139)
        rgb=(0.66151, 0.037182, 0.037182)
        rgb=(0.65328, 0.035225, 0.035225)
        rgb=(0.64503, 0.033268, 0.033268)
        rgb=(0.63674, 0.031311, 0.031311)
        rgb=(0.62842, 0.029354, 0.029354)
        rgb=(0.62008, 0.027397, 0.027397)
        rgb=(0.6117, 0.02544, 0.02544)
        rgb=(0.60329, 0.023483, 0.023483)
        rgb=(0.59485, 0.021526, 0.021526)
        rgb=(0.58638, 0.019569, 0.019569)
        rgb=(0.57788, 0.017613, 0.017613)
        rgb=(0.56935, 0.015656, 0.015656)
        rgb=(0.56079, 0.013699, 0.013699)
        rgb=(0.5522, 0.011742, 0.011742)
        rgb=(0.54357, 0.0097847, 0.0097847)
        rgb=(0.53492, 0.0078278, 0.0078278)
        rgb=(0.52624, 0.0058708, 0.0058708)
        rgb=(0.51752, 0.0039139, 0.0039139)
        rgb=(0.50878, 0.0019569, 0.0019569)
        rgb=(0.5, 0, 0)
    }
}

\maketitle
\thispagestyle{empty}
\pagestyle{empty}

\begin{abstract}
The distributed nature of cellular networks is one of the main enablers for \gls{isac}.
For target positioning and tracking, making use of bistatic measurements is non-trivial due to their non-linear relationship with Cartesian coordinates.
Most of the literature proposes geometric-based methods to determine the target's location by solving a well-defined set of equations stemming from the available measurements.
The error covariance to be used for Bayesian tracking is then derived from local Taylor expansions.
In our work  we  adaptively fuse any subset of bistatic measurements using a \gls{ml} framework, allowing to incorporate every possible combination of available measurements, i.e.,  transmitter angle, receiver angle and bistatic range.
Moreover, our \gls{ml} approach is intrinsically  flexible, as it can be extended to fuse an arbitrary number of measurements by multistatic setups.
Finally, we propose both a fixed and dynamic way to compute the covariance matrix for the position error to be fed to Bayesian tracking techniques, like a \acrlong{kf}.
Numerical evaluations with realistic cellular communications parameters at mmWave frequencies show that our proposal outperforms the considered baselines, achieving a location and velocity \acrlong{rmse} of 0.25\,m and 0.83\,m/s, respectively.
\end{abstract}


\acresetall

\blfootnote{This work has been submitted to the IEEE for possible publication. Copyright may be transferred without notice, after which this version may no longer be accessible.}

\glsresetall
\vspace{-0.5mm}

\section{Introduction}\label{sec:introduction}
\Gls{isac} is expected to be one of the main novel features of future cellular communication standards.
It equips the network with the ability to acquire knowledge about its surrounding environment, enabling new services \cite{mandelli_survey}.
The plethora of applications include detection and localization of pedestrians and automated guided vehicles (AGV) in factories, mobile handover prediction, SLAM and digital twins \cite{liu_isac}.

Those use cases require positioning not only of actively transmitting terminals – which has been available since at least 4G \cite{3gpp_36355} – but also in a radar-like, passive manner through backscattering only.
To enable this, the respective measures have to be extracted from the communications signal~\cite{braun_diss} and can then be fused for positioning a target~\cite{bauhofer_multitarget}.
In addition, one would like to have the capability to track objects over time.

While there is some work on tracking for general radar applications, literature on distributed cellular based tracking is sparse.
The addition of sensing on top of legacy communication services imposes constraints by the cellular system, which was not designed with this task in mind.
Resulting limitations, regulations and backwards compatibility with older standards have to be taken into account.
Further, in contrast to typical radar setups, potential targets can appear in close proximity to the system.
For tracking itself, prior art suggests to operate directly with the received signal \cite{davey_track,lu2023ml} leveraging the synergy of detection and tracking. 
However, factors like multiple targets, bistatic operations, complexity, or the large signaling overhead to a central entity render this approach impractical in current systems.
Based on subspace tracking~\cite{yang_past}, the authors of \cite{hailang_pastangles} propose a low complexity method combining \gls{aod} and \gls{aoa} together with ESPRIT.
In contrast to this, the approach of extracting the target measurements in a first step and then applying tracking is largely suggested by radar literature~\cite{lerro1993tracking}.
However, this requires linearizing the non-Cartesian measurements, which is typically done by extended or unscented \glspl{kf}~\cite{crouse2013ukf,coogle2013ukf}.
More recent work~\cite{marom2023range} and an earlier study \cite{bar_monostaticlongrange} suggest that a linear \gls{kf}, fed with measurements fused to location estimates, results in better performance, known as converted measurement \gls{kf}. 
The necessary measurement error covariance matrix of such an estimate can be approximated by linearization through a low-order Taylor series.

In this work, we leverage an \gls{ml} approach for probabilistically fused position estimates.
We derive the corresponding covariance dynamically from the fusion Hessian matrix.
Both are then fed to a linear \gls{kf}.
This approach can be seen as an extension to the state of the art, with equal performance but more versatile applications, as it can fuse an arbitrary number of measurements.
A second, less complex but even more powerful approach uses the same \gls{ml} location estimate with a position-independent fixed covariance based on prior calibration with the expected \gls{rmse}.
Their main advantage is easy expandability to multistatic scenarios in which measurements from different sensor pairs are combined.
In our 2D campus network simulations, we use bistatic range, \gls{aod}, and \gls{aoa} for a single target tracking task. 
Our main contributions are:
\begin{enumerate}
    \item We derive an \gls{ml} approach to fuse the bistatic information (range, transmit and receive angle) or subsets thereof to localize a target.
    \item We show the similarities of our \gls{ml} approach and state of the art, and the superior performance of the fixed covariance approach, which are both more versatile than the baseline and enhance the performance of \glspl{kf}.
    \item We compare our proposals to the baseline technique with numerical evaluations based on realistic cellular system parametrization and complex target trajectories.
\end{enumerate}

\section{System Model}\label{sec:model}
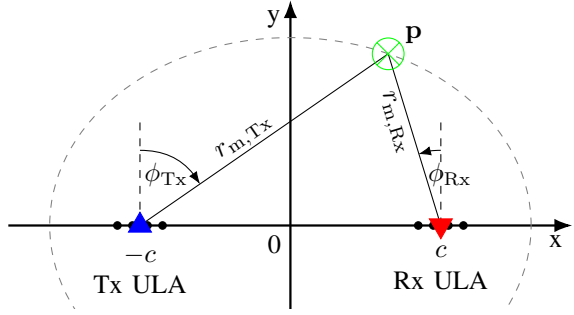
\begin{figure}[t]
    \centering
    \begin{tikzpicture}[x=1mm,y=1mm]
    \def\fwidth{75}

    \clip (0,-11) rectangle (\fwidth,30);

    \draw[-{Latex}, line width=0.3mm] (0,0) -- node[at end,below, anchor=north east] {x} (\fwidth,0);
    \draw[-{Latex}, line width=0.3mm] (\fwidth/2,-30) -- node[at end,left, anchor=north east] {y} (\fwidth/2,30);
    
    \draw[dashed, gray] (\fwidth/2,0) ellipse (32 and 25);
    
    \draw[] (\fwidth/2-20,0) -- node[midway, above, sloped, anchor=south, xshift=-2mm, yshift=-0.9mm] {$r_\mathrm{m,\acrshort{tx}}$} (\fwidth/2+13,22.84);
    \draw (\fwidth/2+20,0) -- node[midway, left, sloped, anchor=north, xshift=-3mm, yshift=0.5mm] {$r_\mathrm{m,\acrshort{rx}}$} (\fwidth/2+13,22.84);

    \node[mark size=6, color=green, rotate=180] at (\fwidth/2+13,22.84) {\pgfuseplotmark{otimes}};
    
    \draw[dashed, black] (\fwidth/2-20,0) -- (\fwidth/2-20,14.5);
    \draw [-{Latex}, black] (\fwidth/2-20,10) to [out=0,in=120] (\fwidth/2-12,5.5);
    \draw[dashed, black] (\fwidth/2+20,0) -- (\fwidth/2+20,14.5);
    \draw [-{Latex}, black] (\fwidth/2+20,10) to [out=180,in=20] (\fwidth/2+17,9.5);
    
    \node[anchor=north east] at (\fwidth/2,0) {0};
    \node[anchor=north, align=center] at (\fwidth/2-20,-1.5) {$-c$\\ \acrshort{tx} \acrshort{ula}};
    \node[anchor=north, align=center] at (\fwidth/2+20,-1.5) {$c$\\ \acrshort{rx} \acrshort{ula}};
    \node[anchor=south west] at (\fwidth/2+14,23) {$\vect{p}$};
    \node[anchor=south] at (\fwidth/2-16.5,4) {$\phi_\mathrm{\acrshort{tx}}$};
    \node[anchor=south] at (\fwidth/2+21.2,4) {$\phi_\mathrm{\acrshort{rx}}$};
    
    \filldraw [draw=black] (\fwidth/2-20-3,0) circle (0.5);
    \filldraw [draw=black] (\fwidth/2-20-1,0) circle (0.5);
    \filldraw [draw=black] (\fwidth/2-20+1,0) circle (0.5);
    \filldraw [draw=black] (\fwidth/2-20+3,0) circle (0.5);
    \node[mark size=5, color=blue, rotate=0] at (\fwidth/2-20,0) {\pgfuseplotmark{triangle*}};
    \filldraw [draw=black] (\fwidth/2+20-3,0) circle (0.5);
    \filldraw [draw=black] (\fwidth/2+20-1,0) circle (0.5);
    \filldraw [draw=black] (\fwidth/2+20+1,0) circle (0.5);
    \filldraw [draw=black] (\fwidth/2+20+3,0) circle (0.5);
    \node[mark size=5, color=red, rotate=180] at (\fwidth/2+20,0) {\pgfuseplotmark{triangle*}};

\end{tikzpicture}
    \caption{Schematic of a bistatic \acrshort{isac} system.
    \acrshort{tx} and \acrshort{rx} \acrshort{ula} are placed at $\pm c$ from the origin in x-direction.
    The target at location $\vect{p}$ is marked with its geometrically derived quantities $r_\mathrm{b}$, $\phi_\mathrm{\acrshort{tx}}$, $\phi_\mathrm{\acrshort{rx}}$, for bistatic range, \acrshort{aod} at \gls{tx}, and \acrshort{aoa} at \gls{rx}, respectively.}
    \label{fig:system_schematic}
\end{figure}

We consider the bistatic two-dimensional system depicted in Fig.~\ref{fig:system_schematic}.
The \gls{tx} and \gls{rx} \glspl{ula} are positioned on the x-axis at $\vect{p}_\mathrm{\acrshort{tx}}=[-c,0]^\intercal$, $\vect{p}_\mathrm{\acrshort{rx}}=[+c,0]^\intercal$, respectively.
From bistatic geometry, the baseline is therefore $2c$, corresponding to half the linear eccentricity for the isorange ellipse with \gls{tx} and \gls{rx} in its foci.
We describe an impulsively scattering target by its state $\boldsymbol{\theta} = [\vect{p},\vect{v}]^\intercal$, with location $\vect{p}=[p_\mathrm{x},p_\mathrm{y}]^\intercal$ and velocity $\vect{v}=[v_\mathrm{x},v_\mathrm{y}]^\intercal$.
We define the bistatic range $r_\mathrm{b}=r_\mathrm{m,\acrshort{tx}}+r_\mathrm{m,\acrshort{rx}}$, \gls{aod} $\phi_\mathrm{\acrshort{tx}}$, and \gls{aoa} $\phi_\mathrm{\acrshort{rx}}$ as
\begin{align}
    r_{\mathrm{b}}\left(\vect{p}\right)
    &= \Vert \vect{p} - \vect{p}_\mathrm{\acrshort{tx}} \Vert_2 + \Vert \vect{p} - \vect{p}_\mathrm{\acrshort{rx}} \Vert_2 \label{eq:range} \\
    &= \sqrt{\left(p_\mathrm{x}+c\right)^2+p_\mathrm{y}^2} + \sqrt{\left(p_\mathrm{x}-c\right)^2+p_\mathrm{y}^2} \, ,\nonumber \\
    \phi_\mathrm{\acrshort{tx}}\left(\vect{p}\right)
    &= -\arctan\left(\frac{p_{\mathrm{x}} + c}{\vert p_{\mathrm{y}} \vert}\right) \, ,\label{eq:angle_tx} \\
    \phi_\mathrm{\acrshort{rx}}\left(\vect{p}\right)
    &= -\arctan\left(\frac{p_{\mathrm{x}} - c}{\vert p_{\mathrm{y}} \vert}\right) \, .
    \label{eq:angle_rx}
\end{align}
The measured angles at the arrays are expressed using the \gls{naf}~\cite{mandelli_sara} given as
\begin{equation}
    \eta\left(\phi\right) = \frac{d}{\lambda_\mathrm{c}} \sin\left(\phi\right) \, ,
    \label{eq:naf}
\end{equation}
with \gls{ula} element distance $d$ and carrier wavelength $\lambda_\mathrm{c}$.\\
The bistatic setup acquires range and angular measurements with \gls{awgn} denoted as  $\alpha_\mathrm{r} \sim \mathcal{N}\left(0,\sigma^2_\mathrm{r}\right)$, $\alpha_\mathrm{\eta_\mathrm{\acrshort{tx}}}, \alpha_\mathrm{\eta_\mathrm{\acrshort{rx}}} \sim \mathcal{N}\left(0,\sigma^2_\mathrm{\eta}\right)$, respectively.
This yields the measurements
\begin{align}
    \tilde{r}_{\mathrm{b}} &= r_{\mathrm{b}} + \alpha_\mathrm{r} \label{eq:noisy1}\\
    \tilde{\eta}_\mathrm{\acrshort{tx}} &= \eta\left(\phi_\mathrm{\acrshort{tx}}\right) + \alpha_\mathrm{\eta_\mathrm{\acrshort{tx}}} \, , \label{eq:noisy2}\\
    \tilde{\eta}_\mathrm{\acrshort{rx}} &= \eta\left(\phi_\mathrm{\acrshort{rx}}\right) + \alpha_\mathrm{\eta_\mathrm{\acrshort{rx}}} \, , \label{eq:noisy3}
\end{align}
where the dependency on $\vect{p}$ has been dropped for readability.

\section{Bistatic Positioning}\label{sec:positioning}
In this section we want to estimate the Cartesian location of a target from the available measurements \mbox{\eqref{eq:noisy1}, \eqref{eq:noisy3}} in the system.
A single measurement will lead to an underdetermined system, from which an unambiguous target location cannot be deduced.
For exactly two measurements a precise location can be derived with prior art geometric-based approaches, which are described in the first subsection.
The second subsection introduces our probabilistic approach, which can even handle an overdetermined system where all three measures are available.
With both approaches we describe also how to estimate the location error's covariance matrix, which indicates a confidence in the derived position to be used for target tracking with Bayesian tracking techniques.

\subsection{Geometric Positioning}
In geometric positioning, exactly two values among range, \gls{aod}, and \gls{aoa} need to be combined to determine the target location.
Combining the two angles, \gls{aod} and \gls{aoa}, yields the following conversion function to Cartesian coordinates
\begin{equation}
    \vect{p}\left(\eta_\mathrm{\acrshort{tx}},\eta_\mathrm{\acrshort{rx}}\right)
    = \begin{bmatrix}
        -c\frac{\tan\left(\phi\left(\eta=\eta_\mathrm{\acrshort{rx}}\right)\right)+\tan\left(\phi\left(\eta=\eta_\mathrm{\acrshort{tx}}\right)\right)}{\tan\left(\phi\left(\eta=\eta_\mathrm{\acrshort{rx}}\right)\right)-\tan\left(\phi\left(\eta=\eta_\mathrm{\acrshort{tx}}\right)\right)} \\
        \frac{2c}{\tan\left(\phi\left(\eta=\eta_\mathrm{\acrshort{rx}}\right)\right)-\tan\left(\phi\left(\eta=\eta_\mathrm{\acrshort{tx}}\right)\right)} \\
    \end{bmatrix} \, ,
    \label{eq:locest_aa}
\end{equation}
computed by inverting~\eqref{eq:naf}.
The other combination of a single angle, \gls{aod} or \gls{aoa}, with bistatic range requires a conversion to the respective \gls{ula}-centric monostatic range
\begin{equation}
    r_\mathrm{m}\left(\phi',r_\mathrm{b}\right) = \frac{r_\mathrm{b}^2 - 4c^2}{2\left(r_\mathrm{b} - 2c\sin\left(\phi'\right)\right)} \, ,
    \label{eq:locest_ar_r}
\end{equation}
with $\phi'=\pi+\phi\left({\eta}_\mathrm{\acrshort{tx}}\right)$ or $\phi'=\phi\left({\eta}_\mathrm{\acrshort{rx}}\right)$ for \gls{tx} or \gls{rx} measurement, respectively.
The final Cartesian location is obtained by a coordinate transform and translation
\begin{equation}
    \vect{p}\left(r_\mathrm{m},\eta,c\right)
    = \begin{bmatrix}
        -r_\mathrm{m} \sin\left(\phi\left(\eta\right)\right) \mp c\\
        r_\mathrm{m} \cos\left(\phi\left(\eta\right)\right)
    \end{bmatrix} \, .
    \label{eq:locest_ar_p}
\end{equation}
For the respective estimation covariance matrix, we employ the first-order Taylor series expansion approach by \cite{bar_monostaticlongrange}.
The components of the estimated location $\vect{\hat{p}}=\vect{p}(\tilde{\vect{m}})$ can be deduced from the noisy measurements vector $\vect{\tilde{m}}$ and the proper conversion function components $p_i(\cdot)$ among \eqref{eq:locest_aa} or \eqref{eq:locest_ar_r}.
Accordingly, we write the Taylor expansion of $\vect{p}(\tilde{\vect{m}})$ around ${\tilde{\vect{m}}}$, keeping only the first-order components as follows
\begin{equation}
        \hat{p}_i = p_i\left(\vect{\tilde{m}}\right) = p_i\left(\vect{m}\right) + \sum_{j=1}^2 \frac{\partial p_i\left(\vect{m}\right)}{\partial m_j} \alpha_j \, .
\end{equation}
Assuming zero-mean \gls{awgn}, we have $\mathbb{E}\left[\alpha_j\right] = 0$, thus  $\mathbb{E}\left[\hat{p}_i\right] = p_i\left(\vect{m}\right)$ and therefore variance $\mathbb{E}\left[|\alpha_j|^2\right] = \sigma_j^2$.
From this, the elements of the converted error covariance matrix, here $\matr{\hat{C}}_\mathrm{Geo}^{2\times2}\left(\vect{\tilde{m}}\right)$, approximating the confidence from measurement to Cartesian space, are
\begin{align}
    \left[\matr{C}_\mathrm{Geo}\left(\vect{m}\right)\right]_{pq}
    &= \mathbb{E}\left[ \sum_{j=1}^2 \frac{\partial p_p\left(\vect{m}\right)}{\partial m_j} \alpha_j \sum_{j=1}^2 \frac{\partial p_q\left(\vect{m}\right)}{\partial m_j} \alpha_j \right] \nonumber \\
    \approx \left[\matr{\hat{C}}_\mathrm{Geo}\left(\vect{\tilde{m}}\right)\right]_{pq}
    &= \sum_{j=1}^2 \frac{\partial p_p\left(\vect{\tilde{m}}\right)}{\partial m_j} \frac{\partial p_q\left(\vect{\tilde{m}}\right)}{\partial m_j} \sigma_j^2
    \label{eq:ecov_geo} \, .
\end{align}
Note that in~\cite{bar_monostaticlongrange} the Taylor expansion was considered up to the second order, since it was proven to be necessary at high distance from \gls{tx} and \gls{rx}.
In our study, for the sake of simplicity and due to the short range applications we target, we limit ourselves to the first order.

\subsection{Maximum Likelihood Positioning}
In this subsection we derive how to fuse the available measurements and estimate the target location with a \gls{ml} approach.
We define the likelihood function for the target position, given $o$ available noisy observations stacked in $\vect{\tilde{m}}$ as
\begin{align}
    \mathcal{L} \left(\vect{\tilde{m}} \vert \vect{p} \right)
    = \frac{\exp{\left(-\frac{1}{2}\left(\vect{f}\left(\vect{p}\right)-\vect{\tilde{m}} \right)^\intercal \matr{C}^{-1} \left(\vect{f}\left(\vect{p}\right)-\vect{\tilde{m}} \right)\right)}}{\sqrt{\left(2\pi\right)^o \det\left(\matr{C}\right)}} \, ,
    \label{eq:likelihood}
\end{align}
where the basis is the multivariate normal distribution due to the assumed \gls{awgn} in the measurements, according to \eqref{eq:noisy1} -- \eqref{eq:noisy3}.
The variances of the available measurements form the diagonal covariance matrix $\matr{C}$.
The vector-function $\vect{f}\left(\vect{p}\right)$ converts a candidate location $\vect{p}$ into the space of available measurements according to \eqref{eq:range} -- \eqref{eq:naf}.
The \gls{ml} estimate is then 
\begin{align}
    \label{eq:ml_estimate}
    \hat{\vect{p}}_\mathrm{\acrshort{ml}}
    &= \argmax_\vect{p} \left( \ln \left( \mathcal{L} \left(\vect{\tilde{m}} \vert \vect{p} \right)\right)\right) \nonumber \\
    &= \argmin_\vect{p} \left( \left[\vect{f}\left(\vect{p}\right) - \vect{\tilde{m}}\right]^\intercal \matr{C}^{-1} \left[\vect{f}\left(\vect{p}\right) - \vect{\tilde{m}}\right] \right) \, .
\end{align}
The error covariance matrix of the location estimate to be fed to Bayesian tracking algorithms is computed as the inverted negative Hessian matrix $\matr{\hat{C}}_\mathrm{\acrshort{ml}} = -\matr{\hat{H}}^{-1}_\mathrm{\acrshort{ml}}$ of the log-likelihood function~\cite{thacker1989role}
\begin{align}
    \left[\matr{\hat{H}}_\mathrm{\acrshort{ml}}\right]_{pq}
    = \frac{\partial^2 \ln \left( \mathcal{L} \left(\vect{\tilde{m}} \vert \vect{p} \right)\right)}{\partial \tilde{m}_p \partial \tilde{m}_q} \, .
    \label{eq:ecov_ml}
\end{align}
Note that this probabilistic approach can estimate -- starting from two different measurement types -- also overdetermined systems with an arbitrary number of measurements.
This can also be leveraged for fusing measurements from multistatic setups with, for example, multiple receivers, by selecting the vector-conversion function $\vect{f}\left(\vect{p}\right)$ appropriately.

As an alternative, one can assume a fixed hand-tuned location error covariance matrix. In our numerical study we consider a diagonal matrix, i.e., assume no correlation between x- and y-coordinate
\begin{align}
    \matr{\hat{C}}_\mathrm{fix} =
    \diag\left(\sigma^2_\mathrm{x}, \sigma^2_\mathrm{y}\right) =
    \begin{bmatrix}
        \sigma^2_\mathrm{x} & 0 \\
        0 & \sigma^2_\mathrm{y} \\
    \end{bmatrix} \, .
    \label{eq:ecov_fix}
\end{align}
This ``fixed'' approach is computationally simple, not requiring any matrix inversion upon each \gls{ml} location estimation.

\section{Kalman Filter based Tracking}\label{sec:tracker}
Measurements acquired over time, while the target moves, can be fused with Bayesian tracking techniques.
In this work we input the location -- and corresponding error covariance estimates -- to a \gls{kf}.
Its linear dynamic system model is
\begin{equation}
    \boldsymbol{\theta}[n] = \matr{A} \cdot \boldsymbol{\theta}[n-1] + \boldsymbol{\beta}[n]
    \label{eq:tracker:mod:proc}
\end{equation}
with state $\boldsymbol{\theta} = \left[p_\mathrm{x},p_\mathrm{y},v_\mathrm{x},v_\mathrm{y}\right]^\intercal$ for each discrete time instance~$n$.
We assume a constant velocity physical prediction model
\begin{equation}
    \matr{A} =
    \begin{bmatrix}
        1 & 0 & \Delta t & 0 \\
        0 & 1 & 0 & \Delta t \\
        0 & 0 & 1 & 0 \\
        0 & 0 & 0 & 1 \\
    \end{bmatrix} \, .
    \label{eq:tracker:mod:A}
\end{equation}
Given the sufficiently high update rate likely to be available in \gls{isac} systems~\cite{wild20236g}, combined with limited acceleration, the process noise realizations $\vect{\beta}[n]$ remain small, justifying the adoption of the constant velocity model~\eqref{eq:tracker:mod:proc}, \eqref{eq:tracker:mod:A}.
The measurement model includes the location estimates $\vect{\hat{p}}$ from (\ref{eq:locest_aa}), (\ref{eq:locest_ar_r}), or (\ref{eq:ml_estimate}), with the error covariance $\matr{\hat{C}}$ either fixed (\ref{eq:ecov_fix}) or dynamically computed from (\ref{eq:ecov_geo}) or (\ref{eq:ecov_ml}).
For the well-known \gls{kf} prediction and update equations, refer, e.g., to~\cite{spagnolini2018statistical}.

\section{Results}\label{sec:results}
This section discusses the simulation setup and the results for positioning and tracking based on numerical evaluations.

\subsection{Simulation Setup}
For the bistatic system, we assume that \gls{tx} and \gls{rx} are perfectly synchronized.
Further, we consider a single impulsively scattering target, such that matching targets to measurements is not necessary.
Based on single set of measurements processing, we work with perfect ground truth and refer to \cite{braun_diss,bauhofer_multitarget} for details on how these would be extracted from, e.g., an \gls{ofdm} transmission.
While we limit ourselves to the 2D case, 3D should be a straightforward extension of the respective equations.
We follow \cite{mandelli_survey} for the system parameters, with an update rate of 10\,ms and noise parameters $\sigma_\mathrm{r}=0.15\,\mathrm{m}$ and $\sigma_\mathrm{\eta}=0.022$, equivalent to $4^\circ$ in boresight.
Given our indoor focus, the baseline (i.e., \gls{tx}-\gls{rx} distance) is selected to be $2c=10\,\mathrm{m}$.\\
For angle-only positioning, we discard realizations which would result in a location estimate behind the baseline, i.e., $y<0$.
The sample locations have a minimum distance of 5\,m to the baseline.
For \gls{ml} fusion, estimates are restricted to $y>0$.
The initial guess is located in a random location up to 3\,m from the ground truth location, which in reality can be obtained using the predicted state from the tracker.
For tracking, position estimates are discarded if they are further than 8\,m away from the previously updated state.
Except for tracking based on \gls{aod}+\gls{aoa} position estimates, we have not enabled any filter reset.
For the two angles case a reset is necessary due to the large position estimation variations, i.e., if there was (i) no state update for the last 0.5\,s, or (ii) the state update lies far outside the evaluated area.
For simplicity, the tracker is initialized and reset with the ground truth state and initial error covariance $\diag\left(0.01,0.01,0.01,0.01\right)$.
Tests with initialization based on the first location estimate with zero velocity showed convergence in around $1\,\text{s}$ with negligible \gls{rmse} loss over a 60\,s long track duration.
The Kalman filter process noise is empirically set to $\diag\left(0.3,0.3,0.3,0.3\right)$.
Regarding the measurement covariance matrix to be input to the \gls{kf}, we evaluate different proposals.
All geometric based approaches are computed according to (\ref{eq:ecov_geo}).
For the \gls{ml} localization, we propose two variants.
First, the explicit Hessian according to \eqref{eq:ecov_ml}, derived analytically in Appendix~\ref{app:AnalHessianDeriv}, as we have seen slight performance improvements compared to numerical approximations.
As an alternative, we evaluate the computationally simple ``fixed'' covariance matrix baseline (\ref{eq:ecov_fix}).
In this approach the diagonal variances are the squared \glspl{rmse} of the position estimation evaluation in Tab.~\ref{tab:positioning_performance}.

For positioning evaluation, we draw 5000 samples from each point on a $10 \times 10$ uniform grid, bounded by $x\in[-15,15]\,\text{m}$ and $y\in[5,35]\,\text{m}$.
\begin{figure}[t]
    \centering
    \begin{tikzpicture}
	\begin{axis}[
		width=0.4\textwidth,
		grid=both, grid style={solid,unigrau!20}, axis equal image,
		xmin=-16,xmax=16,
		ymin=0,ymax=26,
		xtick={-15,-10,-5,0,5,10,15},
		ytick={0,5,...,25},
		xlabel={x in m}, ylabel={y in m},
		xlabel near ticks, ylabel near ticks,
		bar width=3mm,
		legend cell align={left},
		legend style={/tikz/every even column/.append style={column sep=0.1cm}},
	 	legend columns=3,
		legend style={at={(1,1.05)},anchor=south east},
		colorbar, colormap name=jet_inue, colorbar style={ylabel={speed in m/s}, at={(1.05,0)}, anchor=south west, width=2mm, yticklabel style={font=\small}, ylabel style={yshift=1mm, font=\small}, ytick={0,0.5,...,3}, yticklabels={0,0.5,...,3}}, point meta min=0, point meta max=3
		]

        \draw [draw=none, pattern=north east lines, pattern color=red!30!white] (-20,-5) rectangle (20,4.9);
        \draw [draw=none, pattern=north east lines, pattern color=red!30!white] (-20,4.9) rectangle (-15.1,25.1);
        \draw [draw=none, pattern=north east lines, pattern color=red!30!white] (15.1,4.9) rectangle (20,25.1);
        \draw [draw=none, pattern=north east lines, pattern color=red!30!white] (-20,25.1) rectangle (20,30);

        \addplot[mesh, point meta=explicit, line width=2pt] table[x=x, y=y, meta=v, col sep=comma]{./tikz/data/tracking/trajectories/trajectory_sample_8.dat};
        \draw [draw=black] (-10.301,14.398) circle (0.5);
        \draw [draw=black] (7.220,23.335) circle (0.5);
        
        \addplot[mesh, point meta=explicit, line width=2pt] table[x=x, y=y, meta=v, col sep=comma]{./tikz/data/tracking/trajectories/trajectory_sample_28.dat};
        \draw [draw=black] (9.376,5.696) circle (0.5);
        \draw [draw=black] (-12.703,14.322) circle (0.5);
        
        \addplot[only marks, mark=triangle*, mark size=3.5pt, every mark/.append style={rotate=0}, color=blue] coordinates {(-5,0)};
        \addplot[only marks, mark=triangle*, mark size=3.5pt, every mark/.append style={rotate=180}, color=red] coordinates {(5,0.3)};
        
	\end{axis}
\end{tikzpicture}
    \caption{Two exemplary trajectories with 60\,s duration used for tracker evaluation.}
    \label{fig:samples_trajectories}
\end{figure}
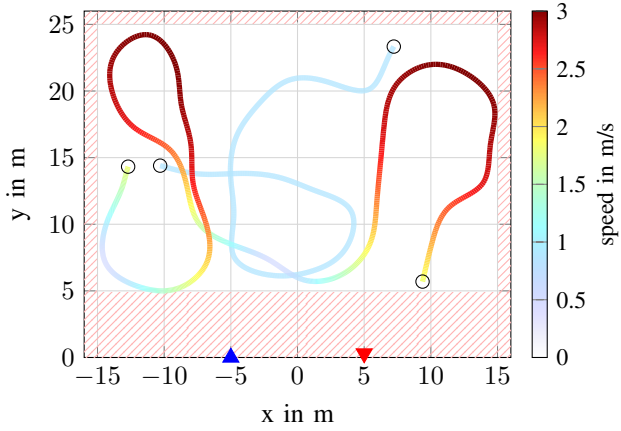
For tracking evaluation, every location estimator is run on 120 tracks, with 40 trials on each.
Fig.~\ref{fig:samples_trajectories} exemplarily depicts two such trajectories.
The track generation follows a pseudorandom walk starting from a random point in the white area shown in Fig.~\ref{fig:samples_trajectories}.
Random change of directions will then generate a series of points, interpolated with cubic-Bezier curves.
The speed profile is generated according to a sinusoidal waveform of random initial phase, period lengths of $\{30,70,100,1\e{8}\}\,\mathrm{s}$, with minimum and maximum values of 0.5 and 3 m/s, respectively, for a duration of 60\,s.

Evaluation is based on the mean absolute error and \gls{rmse} between estimates and ground truth
\begin{equation}
    \mathrm{\acrshort{rmse}} = \sqrt{\mathbb{E}[||\vect{x} - \vect{ \hat{x}}||_2^2]} \; ,
\end{equation}
where $\vect{x}$ is either location $\vect{p}$ or velocity $\vect{v}$.

\begin{figure}[t]
    \centering
    \begin{tikzpicture}
    \def\pwidth{0.27\textwidth}
    \centering
    \begin{groupplot}[
        group style={group name=simulationsamples, group size=2 by 2, vertical sep=10mm, horizontal sep=13mm, x descriptions at=edge bottom, y descriptions at=edge left},
        width=\pwidth,
        grid style={solid,unigrau!20}, grid=both,
		]

        \nextgroupplot[
            height=\pwidth/2, yshift=10mm,
            xmin=-0.5, xmax=0.5, xlabel={angle $\eta$ in NAF}, xlabel near ticks, xtick={-0.5,-0.25,...,0.5}, xticklabels={-0.5,-0.25,...,0.5}, xlabel style={yshift=1.5mm}, minor x tick num=1,
            ymin=0, ymax=13, ytick=\empty,
            ]
            
            \addplot[solid, color=black, name path=d1] table[x=angle, y expr={+\thisrow{d1}}, col sep=comma]{./tikz/data/location_estimation/histogram_naf.dat};
            \addplot[solid, color=black, name path=d2] table[x=angle, y expr={+\thisrow{d2}}, col sep=comma]{./tikz/data/location_estimation/histogram_naf.dat};
            \addplot[solid, color=black, name path=d3] table[x=angle, y expr={+\thisrow{d3}}, col sep=comma]{./tikz/data/location_estimation/histogram_naf.dat};

            \addplot[name path=floor, draw=none] coordinates {(-0.5,0) (0.5,0)};
            \addplot[color=blue!100!white] fill between[of=d1 and floor, soft clip={domain=-0.125:0.125}];
            \addplot[color=blue!60!white] fill between[of=d2 and floor, soft clip={domain=-0.5:0}];
            \addplot[color=blue!30!white] fill between[of=d3 and floor, soft clip={domain=-0.5:0.5}];
   
            \draw [solid, thick, color=red] (0,0) -- (0,20);
            \draw [solid, thick, color=red] (-0.3,0) -- (-0.3,20);
            \draw [solid, thick, color=red] (-0.48,0) -- (-0.48,20);

        \nextgroupplot[
            axis equal image, axis on top,
            xmin=-16, xmax=16, xlabel={x in m}, xtick={-15,-5,...,15}, xticklabels={-15,-5,...,15}, xlabel style={yshift=1.5mm}, minor x tick num=1,
            ymin=-16, ymax=16, ytick={-15,-5,...,15}, yticklabels={-15,-5,...,15}, ylabel={y in m},
            ylabel style={yshift=-3mm}, minor y tick num=0,
            colorbar, colormap name=jet_inue, colorbar style={ylabel={likelihood}, at={(1.05,0)}, anchor=south west, width=2mm, yticklabel style={font=\small}, ylabel style={yshift=7mm, font=\small}, ytick={1.5,18.5}, tick style={draw opacity=0}, tick style={color=white}, yticklabels={low,high}}, point meta min=0, point meta max=20,
            ]

            \addplot graphics [xmin=-17, xmax=17, ymin=-17, ymax=17]{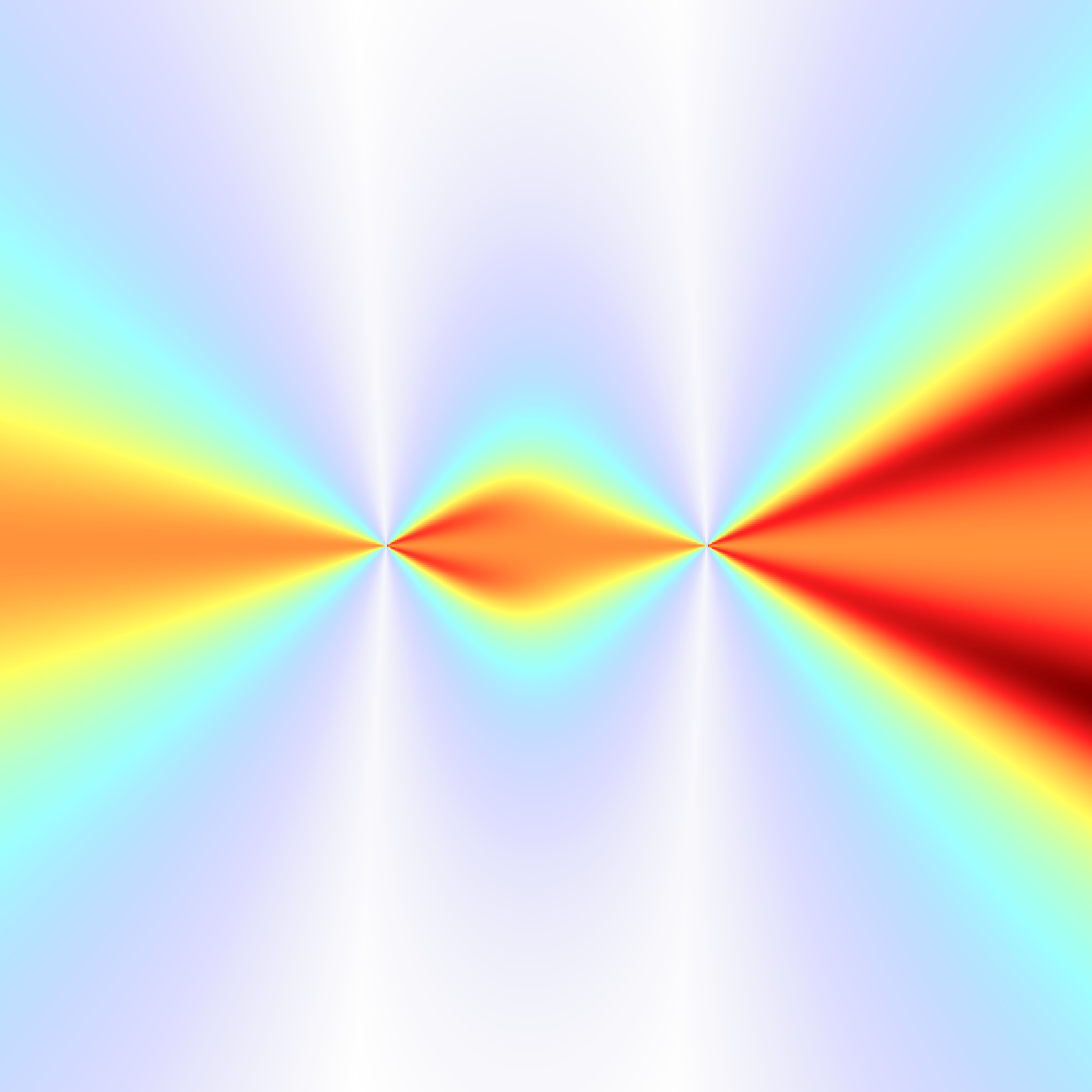};

            \addplot[only marks, mark=triangle*, mark size=3.5pt, every mark/.append style={rotate=0}, color=blue, forget plot] coordinates {(-5,0)};
            \addplot[only marks, mark=triangle*, mark size=3.5pt, every mark/.append style={rotate=180}, color=red, forget plot] coordinates {(5,0.3)};

            \addplot[only marks, mark=+, mark size=4pt, every mark/.append style={rotate=0}, color=uni_mittelblau] coordinates {(10,5)};
            \label{plots:gt_target}
            \addplot[mark=none, uni_mittelblau, dashed, forget plot] coordinates {(5,0) (20,15)};
            \addplot[mark=none, uni_mittelblau, dashed, forget plot] coordinates {(-5,0) (25,10)};

            \addplot[only marks, mark=x, mark size=4pt, every mark/.append style={rotate=0}, color=black] coordinates {(13,3)};
            \label{plots:gt_estimate}
            \addplot[mark=none, black, forget plot] coordinates {(5,0) (20,10)};
            \addplot[mark=none, black] coordinates {(-5,0) (-16,1)};
            \label{plots:est_target_lines}

        \nextgroupplot[
            height=\pwidth/2, yshift=11mm,
            xmin=-1.571, xmax=1.571, xlabel={angle $\phi$ in rad}, xlabel near ticks, xtick={-1.571,-1.178,-0.785,-0.393,0,0.393,0.785,1.178,1.571}, xticklabels={$-\pi/2$,,$-\pi/4$,,0,,$-\pi/4$,,$\pi/2$}, xlabel style={yshift=1.5mm}, minor x tick num=1,
            ymin=0, ymax=7, ytick=\empty,
            ]

            \addplot[solid, color=black, name path=d1] table[x=angle, y=d1, col sep=comma]{./tikz/data/location_estimation/histogram_rad.dat};
            \addplot[solid, color=black, name path=d2] table[x=angle, y=d2, col sep=comma]{./tikz/data/location_estimation/histogram_rad.dat};
            \addplot[solid, color=black, name path=d3] table[x=angle, y=d3, col sep=comma]{./tikz/data/location_estimation/histogram_rad.dat};

            \addplot[name path=floor, draw=none] coordinates {(-1.5,0) (1.5,0)};
            \addplot[color=blue!100!white] fill between[of=d1 and floor, soft clip={domain=-1.6:1.6}];
            \addplot[color=blue!60!white] fill between[of=d2 and floor, soft clip={domain=-1.6:1.6}];
            \addplot[color=blue!30!white] fill between[of=d3 and floor, soft clip={domain=-1.6:1.6}];

            \draw [solid, thick, color=red] (0,0) -- (0,20);
            \draw [solid, thick, color=red] (-0.64,0) -- (-0.64,20);
            \draw [solid, thick, color=red] (-1.287,0) -- (-1.287,20);

	\end{groupplot}
\end{tikzpicture}
    \vspace*{-15mm}
    \caption{\acrshort{ula} and \acrshort{naf} peculiarities.
    \textit{(left)} Histogram for three distinct noisy incident angles with constant variance in \acrshort{naf} and its influence on the radiant domain.
    \textit{(right)} \acrshort{ml} heatmap of an estimated target location (\ref{plots:gt_estimate}), based on noisy \acrshort{aod} and \acrshort{aoa} measurements (\ref{plots:est_target_lines}) from its ground-truth location (\ref{plots:gt_target}).}
    \label{fig:simulation_examples}
\end{figure}
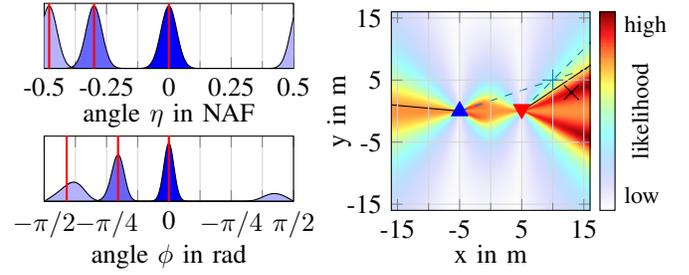

From Fig.~\ref{fig:simulation_examples}, we observe a number of peculiarities when working with \glspl{ula}.
The left figure shows the difference in angle distribution from \gls{naf} to radiant domain.
The upper \gls{naf} image depicts the histogram for three distinct incident angles in red after adding noise with constant variance.
Transforming this into the radiant domain in the lower left using the non-linear trigonometric function in (\ref{eq:naf}) reveals a broadening of the distribution away from boresight.
From this relation, the Gaussian noise distribution also changes its shape.
In both images, the light blue distribution showcases the \gls{ula} property of angular wrapping (i.e., periodicity) around $\pm90^\circ$.
A similar behavior can be observed in the right figure, which depicts the likelihood value of every location given two noisy angle measurements.
Again, the periodic angle behavior can be observed from the transmitter angle wrapped to the other side.
A 1D \gls{ula} cannot distinguish the responses from frontside and backside, resulting in a mirror behavior around $y=0$.
The \gls{ml} estimate follows this property with a relatively high likelihood for locations $x<-5\,\text{m}$.
Note that the constant likelihood on the baseline between \gls{tx} and \gls{rx}, where measured angles do not intersect but overlap with constant range measurements, is consistent with bistatic radar theory.

\subsection{Location Estimation}
\begin{figure}[b!]
    \centering
    \begin{tikzpicture}
    \def\pwidth{0.25\textwidth}
    \centering
    \begin{groupplot}[
        group style={group name=locationsamples, group size=3 by 1, vertical sep=0mm, horizontal sep=2mm},
        width=\pwidth, axis equal image,
        grid=both, grid style={solid,unigrau!20},
        xmin=-19, xmax=19, xlabel={x in m}, xlabel near ticks, xtick={-15,-5,...,15}, xlabel style={yshift=1mm}, minor x tick num=1,
        ymin=0, ymax=37, ytick={5,15,...,35}, yticklabels={,,}, ylabel={},
        ylabel near ticks, ylabel style={yshift=-2mm}, minor y tick num=1,
        legend image post style={scale=3},
		]

        \nextgroupplot[ylabel={y in m}, yticklabels={5,15,...,35}]
        
            \addplot[only marks, mark=x, mark size=1pt, color=uni_mittelblau] table[x=x, y=y, col sep=comma]{./tikz/data/location_estimation/samples/LE_samples_mlArAt.dat};
            \label{plots:ml_AtAr}

            \addplot[only marks, mark=triangle*, mark size=3.5pt, every mark/.append style={rotate=0}, color=blue] coordinates {(-5,0)};
            \addplot[only marks, mark=triangle*, mark size=3.5pt, every mark/.append style={rotate=180}, color=red] coordinates {(5,0.3)};
            
            \addplot[only marks, mark=+, mark size=2pt, color=black] table[x=x, y=y, col sep=comma]{./tikz/data/location_estimation/LE_groundtruth_5x5.dat};

        \nextgroupplot[]

            \addplot[only marks, mark=x, mark size=1pt, color=uni_apfelgruen] table[x=x, y=y, col sep=comma]{./tikz/data/location_estimation/samples/LE_samples_mlArR.dat};
            \label{plots:ml_ArR}

            \addplot[only marks, mark=triangle*, mark size=3.5pt, every mark/.append style={rotate=0}, color=blue] coordinates {(-5,0)};
            \addplot[only marks, mark=triangle*, mark size=3.5pt, every mark/.append style={rotate=180}, color=red] coordinates {(5,0.3)};
            
            \addplot[only marks, mark=+, mark size=2pt, color=black] table[x=x, y=y, col sep=comma]{./tikz/data/location_estimation/LE_groundtruth_5x5.dat};
        
        \nextgroupplot[]

            \addplot[only marks, mark=x, mark size=1pt, color=uni_rot] table[x=x, y=y, col sep=comma]{./tikz/data/location_estimation/samples/LE_samples_mlArAtR.dat};
            \label{plots:ml_AAR}

            \addplot[only marks, mark=triangle*, mark size=3.5pt, every mark/.append style={rotate=0}, color=blue] coordinates {(-5,0)};
            \addplot[only marks, mark=triangle*, mark size=3.5pt, every mark/.append style={rotate=180}, color=red] coordinates {(5,0.3)};
            
            \addplot[only marks, mark=+, mark size=2pt, color=black] table[x=x, y=y, col sep=comma]{./tikz/data/location_estimation/LE_groundtruth_5x5.dat};
            
	\end{groupplot}

    \matrix[matrix of nodes, anchor=south east, draw, inner sep=0.2em, draw] at ([yshift=1.8mm]locationsamples c3r1.north east)
        {
        \ref{plots:ml_AtAr} & \acrshort{aod}+\acrshort{aoa} & [7.3mm]
        \ref{plots:ml_ArR} & \acrshort{aoa}+R & [7.3mm]
        \ref{plots:ml_AAR} & \acrshort{aod}+\acrshort{aoa}+R\\
        };
 
\end{tikzpicture}
 
    \caption{Distribution of location estimates from \acrshort{ml} fusion of different measurements.
    For the sake of overview, only a $5\times5$ sample grid with 154 realizations per point is plotted.}
    \label{fig:simulation_samples_location}
\end{figure}
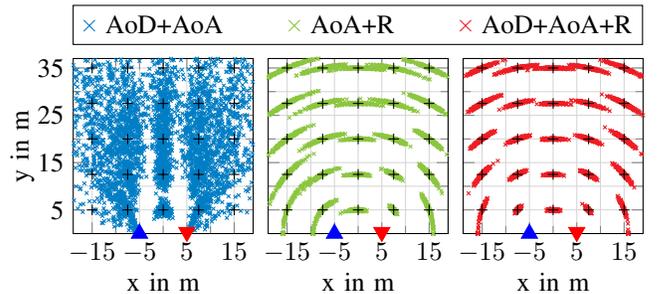

\begin{figure*}[t!]
	\centering
  	\begin{tikzpicture}
    \def\pwidth{0.31\textwidth}
    \centering
    \begin{groupplot}[
        group style={group name=rmseheatmaps, group size=5 by 1, vertical sep=0mm, horizontal sep=4mm},
        width=\pwidth, axis equal image,
        grid=both, grid style={solid,unigrau!20}, axis on top,
        xmin=-16, xmax=16, xlabel={x in m}, xlabel near ticks, xtick={-15,-5,...,15}, xlabel style={yshift=1mm}, minor x tick num=1,
        ymin=0, ymax=36, yticklabels={,,}, ylabel={},
        ylabel near ticks, ylabel style={yshift=-2mm}, minor y tick num=1,
        ]

        \nextgroupplot[ylabel={y in m}, ytick={5,15,...,35}, yticklabels={5,15,...,35}]

            \addplot graphics [xmin=-16, xmax=16, ymin=0, ymax=36]{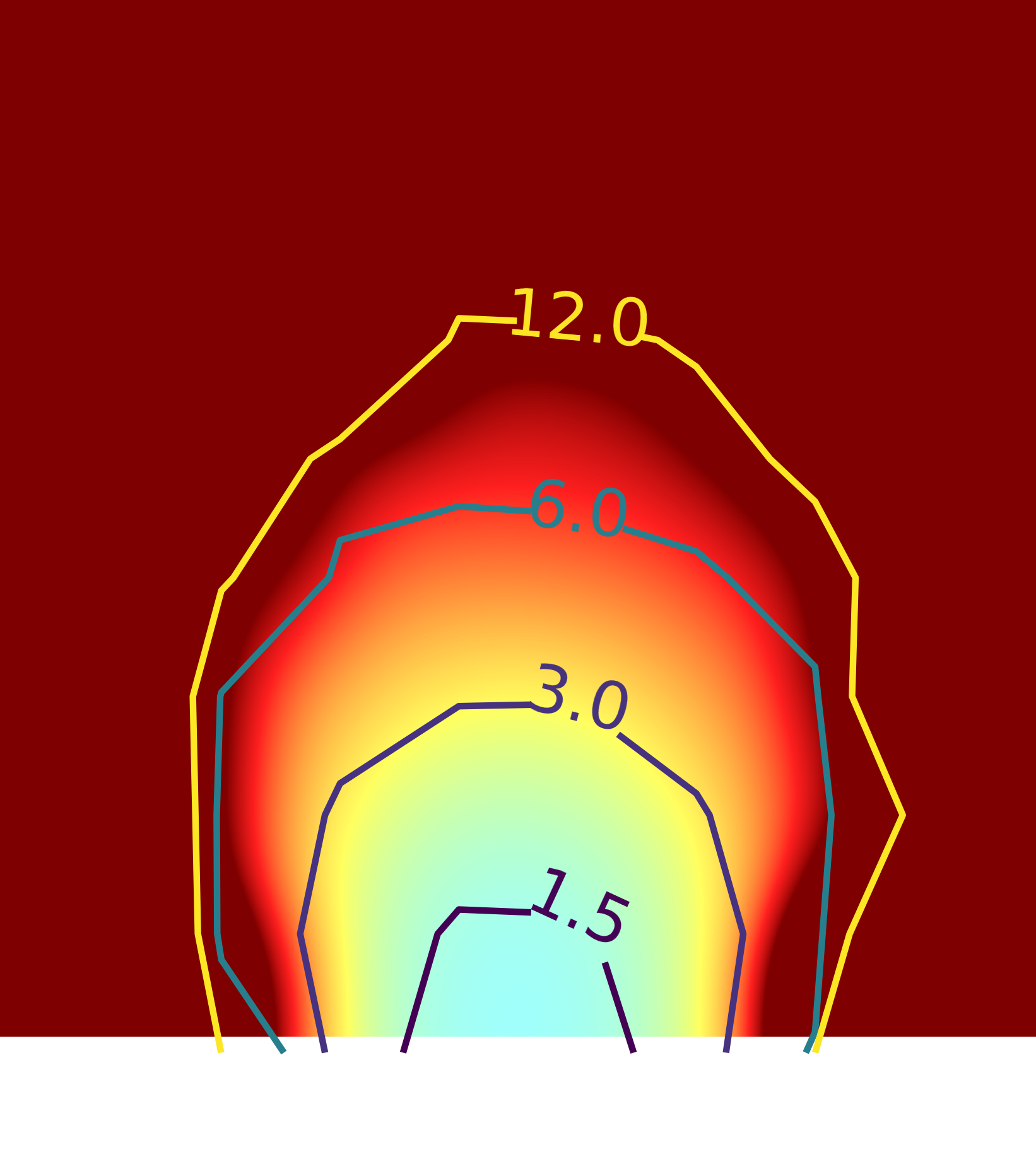};
    
            \draw [draw=none, pattern=north east lines, pattern color=red!30!white] (-20,0) rectangle (20,4.7);
    
            \addplot[only marks, mark=triangle*, mark size=3.5pt, every mark/.append style={rotate=0}, color=blue] coordinates {(-5,0)};
            \addplot[only marks, mark=triangle*, mark size=3.5pt, every mark/.append style={rotate=180}, color=red] coordinates {(5,0.3)};

        \nextgroupplot[]

            \addplot graphics [xmin=-16, xmax=16, ymin=0, ymax=36]{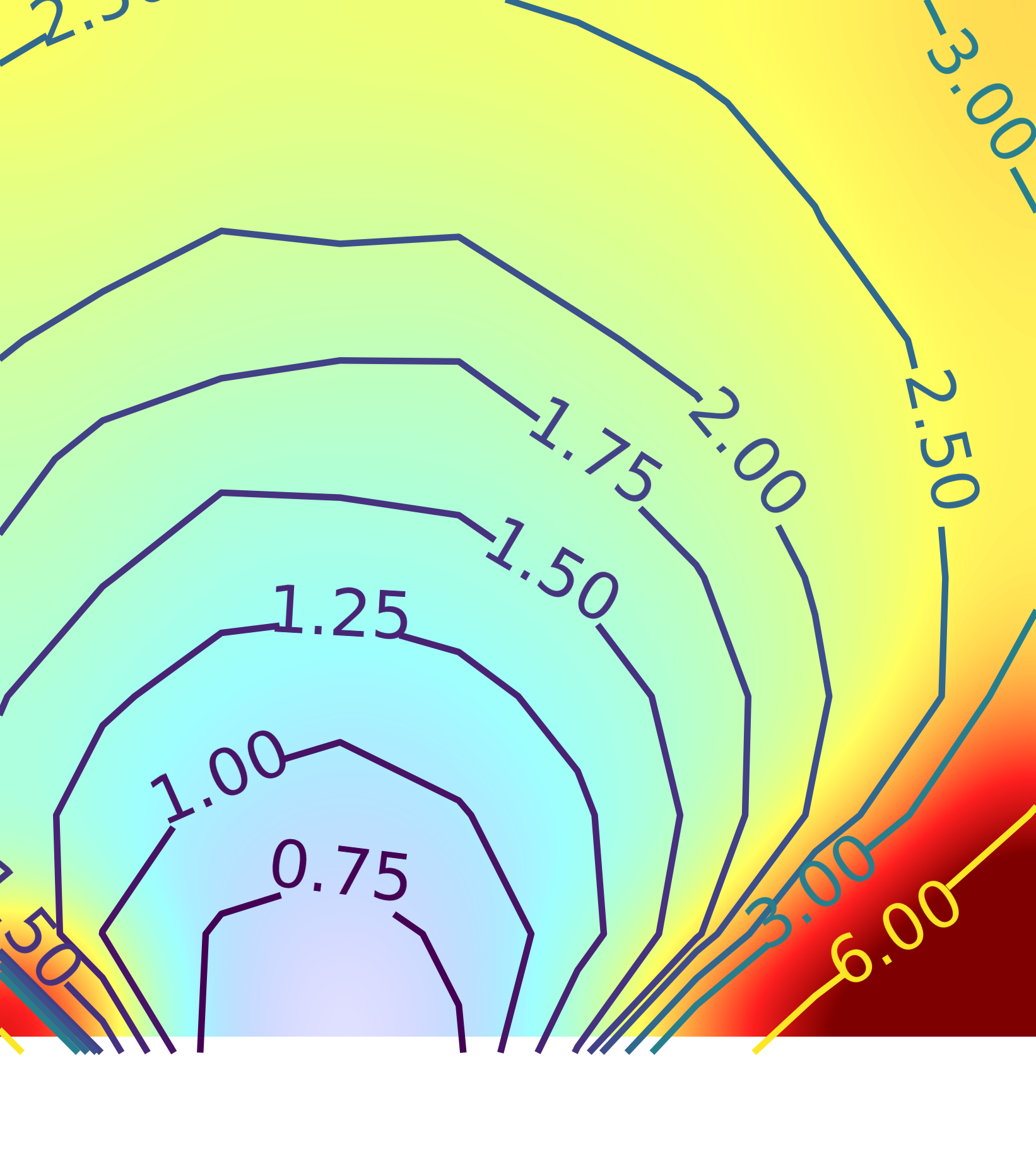};

            \draw [draw=none, pattern=north east lines, pattern color=red!30!white] (-20,0) rectangle (20,4.7);
            
            \addplot[only marks, mark=triangle*, mark size=3.5pt, every mark/.append style={rotate=0}, color=blue] coordinates {(-5,0)};
            \addplot[only marks, mark=triangle*, mark size=3.5pt, every mark/.append style={rotate=180}, color=red] coordinates {(5,0.3)};

        \nextgroupplot[]

            \addplot graphics [xmin=-16, xmax=16, ymin=0, ymax=36]{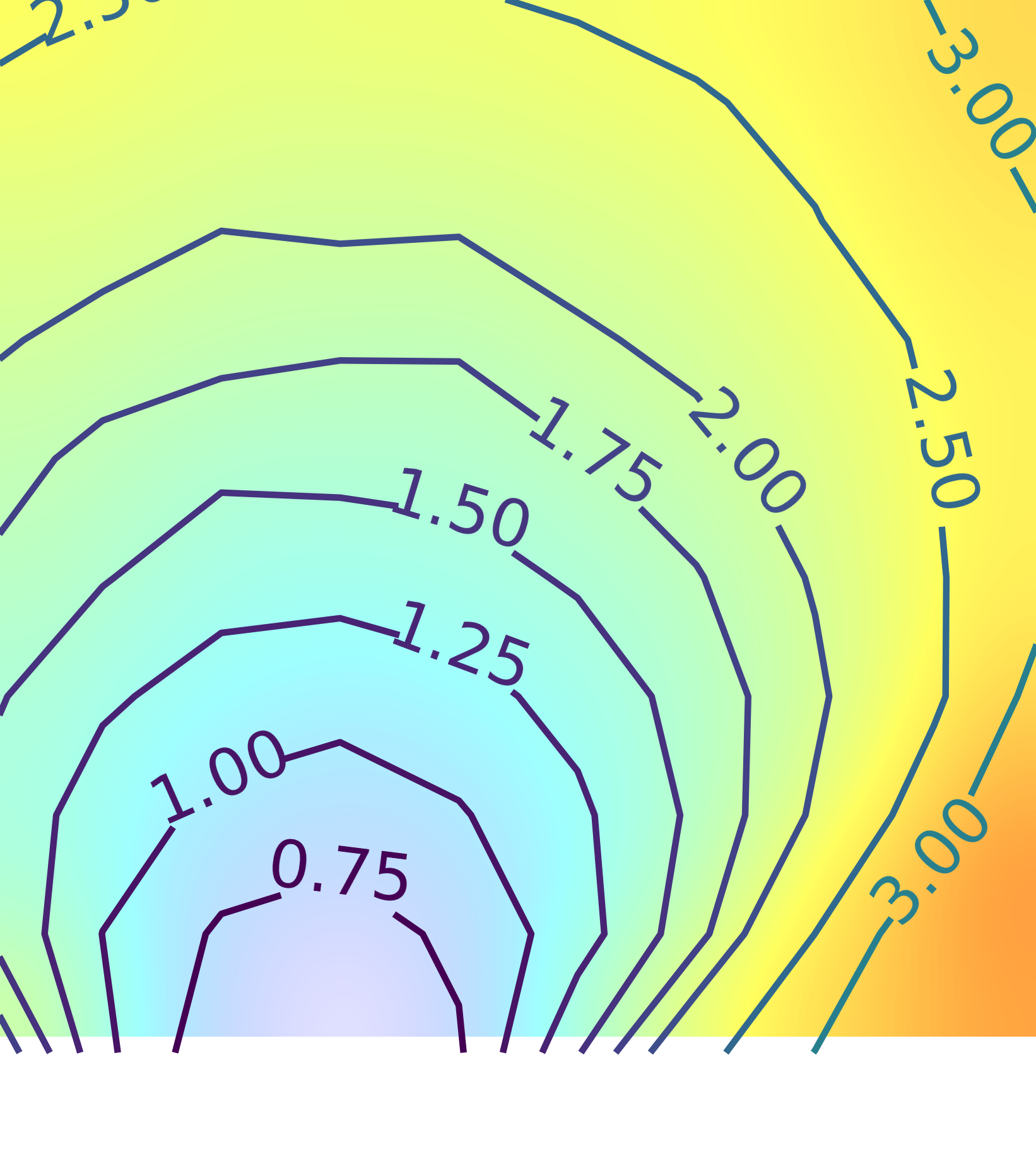};

            \draw [draw=none, pattern=north east lines, pattern color=red!30!white] (-20,0) rectangle (20,4.7);

            \addplot[only marks, mark=triangle*, mark size=3.5pt, every mark/.append style={rotate=0}, color=blue] coordinates {(-5,0)};
            \addplot[only marks, mark=triangle*, mark size=3.5pt, every mark/.append style={rotate=180}, color=red] coordinates {(5,0.3)};

        \nextgroupplot[]

            \addplot graphics [xmin=-16, xmax=16, ymin=0, ymax=36]{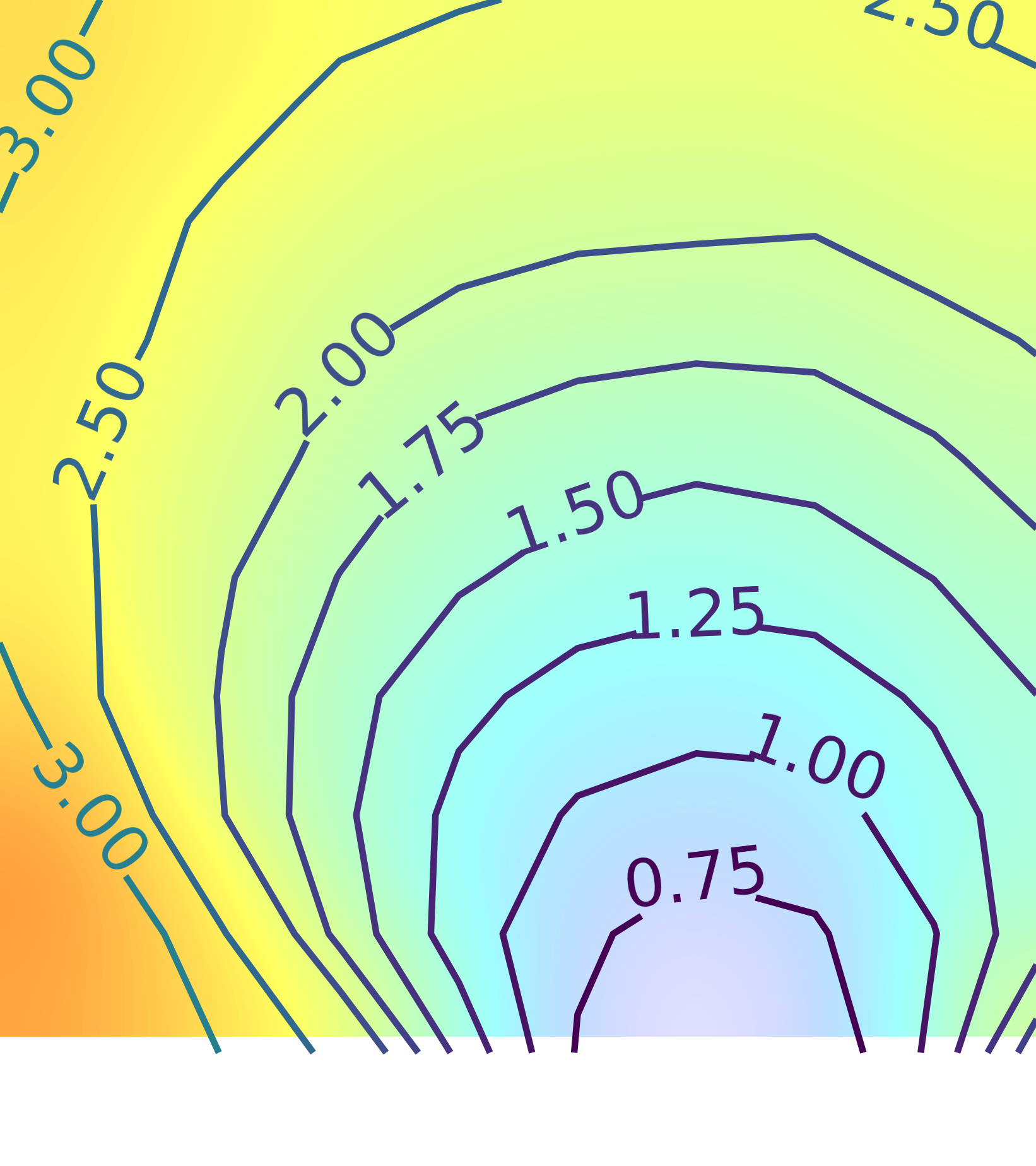};

            \draw [draw=none, pattern=north east lines, pattern color=red!30!white] (-20,0) rectangle (20,4.7);
            
            \addplot[only marks, mark=triangle*, mark size=3.5pt, every mark/.append style={rotate=0}, color=blue] coordinates {(-5,0)};
            \addplot[only marks, mark=triangle*, mark size=3.5pt, every mark/.append style={rotate=180}, color=red] coordinates {(5,0.3)};

        \nextgroupplot[
            colorbar, colormap name=jet_inue, colorbar style={ylabel={\acrshort{rmse} in dB\,m}, at={(1.05,0)}, anchor=south west, yticklabel style={font=\small}, ylabel style={yshift=5mm, font=\small}, ytick={-10,-5,...,20}, width=10mm}, point meta min=-10, point meta max=20,
            ]

            \addplot graphics [xmin=-16, xmax=16, ymin=0, ymax=36]{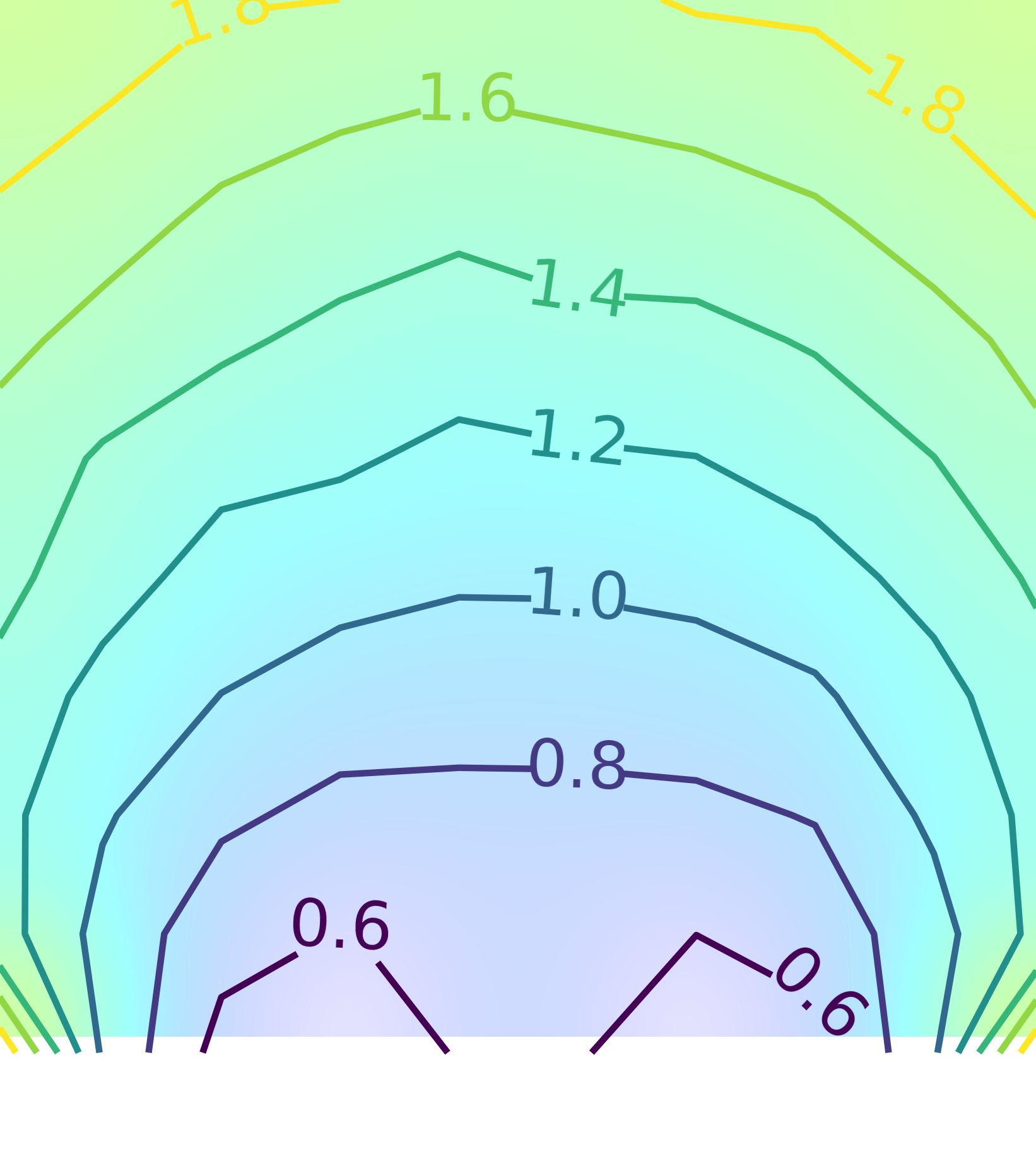};

            \draw [draw=none, pattern=north east lines, pattern color=red!30!white] (-20,0) rectangle (20,4.7);
            
            \addplot[only marks, mark=triangle*, mark size=3.5pt, every mark/.append style={rotate=0}, color=blue] coordinates {(-5,0)};
            \addplot[only marks, mark=triangle*, mark size=3.5pt, every mark/.append style={rotate=180}, color=red] coordinates {(5,0.3)};

	\end{groupplot}
\end{tikzpicture}
	\caption{Exemplary positioning \acrshort{rmse} distribution for a $30\times35$\,m scenario using different approaches and combined measurements.
    (1) \acrshort{ml} \acrshort{aod}+\acrshort{aoa}, (2) geometric \acrshort{aod}+range, (3) \acrshort{ml} \acrshort{aod}+range, (4) \acrshort{ml} \acrshort{aoa}+range, (5) \acrshort{ml} \acrshort{aod}+\acrshort{aoa}+range. The heatmap is based on the $10\times10$ grid with bicubic interpolation between points in decibel and contours in absolute scale.}
	\label{fig:le_heatmap}
\end{figure*}
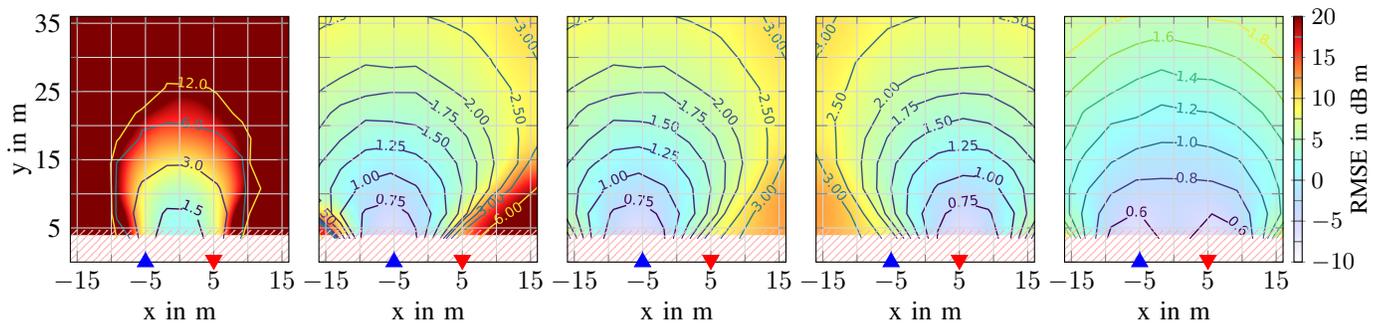

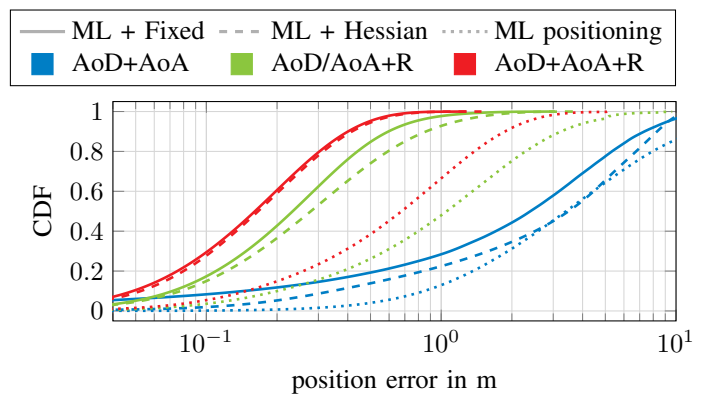
\begin{figure}[!ht]
    \centering
    \begin{tikzpicture}
	\begin{axis}[
		name=cdfs, width=0.5\textwidth, height=4.5cm,
		grid=both, grid style={solid,unigrau!20},
		xmode=log,
		xmin=4e-2,xmax=1e1,
		ymin=-0.05,ymax=1.05,
		ytick={0,0.2,...,1},
		xlabel={position error in m}, ylabel={CDF},
		xlabel near ticks, ylabel near ticks,
		bar width=3mm,
		legend cell align={left},
		legend style={/tikz/every even column/.append style={column sep=0.1cm}},
	 	legend columns=3,
	    legend style={at={(1,1.07)},anchor=south east},
		]

        \addlegendimage{darkgray176, solid, very thick}
        \addlegendentry{ML + Fixed }
        \addlegendimage{darkgray176, dashed, very thick}
        \addlegendentry{ML + Hessian}
        \addlegendimage{darkgray176, dotted, very thick}
        \addlegendentry{\acrshort{ml} positioning}

        \addlegendimage{only marks, mark=square*, uni_mittelblau, mark size=4pt}
        \addlegendentry{\acrshort{aod}+\acrshort{aoa}}
        \addlegendimage{only marks, mark=square*, uni_apfelgruen, mark size=4pt}
        \addlegendentry{\acrshort{aod}/\acrshort{aoa}+R}
        \addlegendimage{only marks, mark=square*, uni_rot, mark size=4pt}
        \addlegendentry{\acrshort{aod}+\acrshort{aoa}+R}

        \addplot[color=uni_mittelblau, solid, line width=1pt] table[x=rmse, y=cdf_percentile, col sep=comma]{./tikz/data/tracking/cdf/Tracker_cdf_fix_aa.dat};
        \addplot[color=uni_apfelgruen, solid, line width=1pt] table[x=rmse, y=cdf_percentile, col sep=comma]{./tikz/data/tracking/cdf/Tracker_cdf_fix_arr.dat};
        \addplot[color=uni_rot, solid, line width=1pt] table[x=rmse, y=cdf_percentile, col sep=comma]{./tikz/data/tracking/cdf/Tracker_cdf_fix_aar.dat};

        \addplot[color=uni_mittelblau, dashed, line width=1pt] table[x=rmse, y=cdf_percentile, col sep=comma]{./tikz/data/tracking/cdf/Tracker_cdf_ml_aa.dat};
        \label{plots:ml_track_AA}
        \addplot[color=uni_apfelgruen, dashed, line width=1pt] table[x=rmse, y=cdf_percentile, col sep=comma]{./tikz/data/tracking/cdf/Tracker_cdf_ml_arr.dat};
        \label{plots:ml_track_AR}
        \addplot[color=uni_rot, dashed, line width=1pt] table[x=rmse, y=cdf_percentile, col sep=comma]{./tikz/data/tracking/cdf/Tracker_cdf_ml_aar.dat};

        \addplot[color=uni_mittelblau, dotted, line width=1pt] table[x=rmse, y=cdf_percentile, col sep=comma]{./tikz/data/location_estimation/cdf/LE_cdf_mlArAt_trackerarea.dat};
        \addplot[color=uni_apfelgruen, dotted, line width=1pt] table[x=rmse, y=cdf_percentile, col sep=comma]{./tikz/data/location_estimation/cdf/LE_cdf_mlArR_trackerarea.dat};
        \addplot[color=uni_rot, dotted, line width=1pt] table[x=rmse, y=cdf_percentile, col sep=comma]{./tikz/data/location_estimation/cdf/LE_cdf_mlArAtR_trackerarea.dat};
	\end{axis}
\end{tikzpicture}
    \vspace*{-3mm}
    \caption{\acrshort{cdf}s of the expected error for positioning only and tracking with \gls{kf} with the Fixed and Hessian approaches.
    The evaluated area is $x \in [-15,15]$\,m and $y \in [-5,25]$\,m.}
    \label{fig:cdfs}
\end{figure}

For location estimation, Fig.~\ref{fig:simulation_samples_location} shows a selection of the distribution of location estimates from different fusion types.
In the cropped left image, we can observe the fanning-out of estimates for the fusion of \gls{aod} and \gls{aoa}, irrespective of the approach.
This results in almost unbounded location estimates for $\eta_\mathrm{\acrshort{tx}} \approx \eta_\mathrm{\acrshort{rx}}$.
In the center image, substituting one angle with bistatic range significantly lowers the sample spread, resulting in a contact lens shaped distribution well-known from bistatic radar.
The estimates are now asymmetrically distributed with lower variance around the observing \gls{ula}.
The same figure but for the geometric positioning approach would reveal angles reappearing on the other side, as described in Fig.~\ref{fig:simulation_examples}.
The right image combines all measurements, which is only possible for the \gls{ml} approach, resulting in a symmetric image with the lowest sample spread.
The conversion of these samples to location \gls{rmse} heatmaps is shown in Fig.~\ref{fig:le_heatmap}.
Geometric and \gls{ml} approach show equal performance for two angles.
For a single angle with range (Figs.~\ref{fig:le_heatmap}.2 and~\ref{fig:le_heatmap}.3) there is a slight improvement for the latter in areas close to the baseline, arising from the described angle continuity.
Figs.~\ref{fig:le_heatmap}.3 and~\ref{fig:le_heatmap}.4 show that a change of measured angle just mirrors the asymmetrical image.
In general, locations closer to the angle observing \gls{ula} yield lower variance due to the low angular spread.
To quantify the performance, Tab.~\ref{tab:positioning_performance} provides the respective \glspl{rmse} averaged over the trajectory area, also used in the fixed covariance baseline tracker.
\begin{table}[H]
	\caption{\gls{ml} position estimation \gls{rmse} performance, average of points within $x \in [-15,15]$\,m and $y \in [-5,25]$\,m.}
	\label{tab:positioning_performance}
	\resizebox{\columnwidth}{!}{%
    
\begin{tabular}{llllll}
    \toprule
    Quantity & \gls{aod}+\gls{aoa} & \gls{aod}+R & \gls{aoa}+R & \gls{aod}+\gls{aoa}+R \\
    \midrule
    x in m & 353.72 & 1.44 & 1.44 & 0.9 \\
    y in m & 90.71 & 0.99 & 0.98 & 0.53 \\
    $\lVert\cdot\rVert_2$ in m & 371.68 & 1.85 & 1.84 & 1.1 \\
    \bottomrule
\end{tabular}

    }
\end{table}
The dotted position error \gls{cdf} curves in Fig.~\ref{fig:cdfs} confirm these results for later comparison with tracking.
We plot \gls{ml} only, with the geometric approach slightly degrading at the high percentiles for fusion of one angle and range.
Here, due to the symmetric sample points, the chosen angle is irrelevant.
Overall, using the overdetermined system -- incorporating a range measurement on top of two angles -- we see a decrease of 46\%, resulting in a mean absolute position error of 2.32\,m for the 95th percentile.
\subsection{Tracking Performance}
Running the tracking algorithms, the solid and dashed lines in Fig.~\ref{fig:cdfs} show the location performance for the assumed fixed covariance matrix and our \gls{ml} approach, respectively.
We omit the curves for state of the art tracking by geometric positioning as they overlap with (\ref{plots:ml_track_AA},\ref{plots:ml_track_AR}).
Overall, tracking improves the positioning error by 57 -- 82\%.
In our scenario, tracking with a position-independent ``Fixed'' covariance matrix does not only outperform the state of the art baseline, but also the Hessian-based approach.
This suggests that the tracker adheres too strongly to the measurements with lower trust in its predicted state.
We observed that the approaches relying on the Hessian and the Taylor approximation used in the geometric results follow this trend.
These capture the strong non-Gaussian/elliptical contact lense shaped distribution away from boresight.
We conclude that using prior information, e.g., through measurements, is beneficial for tracking for this use case and also reduces the computational load.
\begin{table}[b]
	\caption{Tracker \gls{rmse} performance}
	\label{tab:tracker_performance}
    \resizebox{\columnwidth}{!}{%
	
\begin{tabular}{llllll}
    \toprule
    Est. Method & Quantity & \gls{aod}+\gls{aoa} & \gls{aod}+R & \gls{aoa}+R & \gls{aod}+\gls{aoa}+R \\
    \midrule
    \multirow{2}{*}{Geometric} & Location in m & 4.8 & 0.56 & 0.53 & n.a. \\
     & Velocity in m/s & 1.19 & 0.89 & 0.88 & n.a. \\
    \midrule
    \multirow{2}{*}{\acrshort{ml} - Hessian} & Location in m & 4.8 & 0.56 & 0.53 & 0.26 \\
     & Velocity in m/s & 1.19 & 0.89 & 0.88 & 0.81 \\
    \midrule
    \multirow{2}{*}{\acrshort{ml} - Fixed} & Location in m & 4.21 & 0.4 & 0.38 & 0.25 \\
     & Velocity in m/s & 1.63 & 0.9 & 0.9 & 0.83 \\
    \bottomrule
\end{tabular}

    }
\end{table}
Tab.~\ref{tab:tracker_performance} shows the detailed averaged results for tracking.
Coherent with the results on positioning, we see that -- compared to just using angles -- incorporating a range measurement is beneficial.
The combination of our \gls{ml} proposal together with a sample based fixed covariance leads to a more consistent tracker due to a better error covariance estimate.
Especially when combining all measurements and mainly boresight targets, the Hessian-based covariance might be an option with almost no positioning degradation, slightly better velocity estimation and no need for pre-processing/sampling the scenario for the values in Tab.~\ref{tab:positioning_performance}.
Again, this performance is further enhanced by incorporating all available measurements, which is only possible with our ``\gls{ml} - Fixed'' approach, resulting in a location \gls{rmse} down to 0.25\,m.
We observe the same behavior for the velocity estimate, although higher in magnitude.
This relatively high error results from the high accelerations and should be lower in real-world trajectories.

\section{Conclusion}\label{sec:conclusion}
In this paper, we have presented a versatile \gls{ml} approach for the fusion of bistatic measurements into a target location estimate.
We demonstrated its performance with 2D numerical evaluations based on an \gls{isac} system parametrization for an indoor scenario with a single target on complex trajectories.
The combination of our approach with a linear \gls{kf} outperforms current state of the art geometric based techniques with respect to tracking performance by 53\% and 6\%, for a location and velocity \gls{rmse} of 0.25\,m and 0.83\,m/s, respectively.
In addition, we showcased the sub-optimality of approaches relying on Taylor approximation or Hessian to compute the localization error covariance and proposed a superior heuristic with even lower complexity. 
Another benefit of this probabilistic method is the possibility of combining an arbitrary number of combined measurements, enabling multistatic setups with multiple \gls{tx} and/or and \gls{rx}.\\
Incorporating also Doppler information, extension to 3D multi-target capabilities with target-to-measurement association and validation with measurements are topics for future work.

\section*{Acknowledgments}
The authors would like to thank Alexander Felix for his valuable feedback during the development of this work.

This work was developed within the KOMSENS-6G project, partly funded by the German Ministry of Education and Research under grant 16KISK112K.

\begin{appendices}
\appendices
 
\section{Derivations for Analytical Hessian Matrices}\label{app:AnalHessianDeriv}
To derive the Hessian respectively location error covariance matrices, we use (\ref{eq:ecov_ml}) and feed the respective log-likelihood function from~(\ref{eq:likelihood}).
Depending on the fusion type with different $\vect{\tilde{m}}$, Tab.~\ref{tab:analytical_hessians} lists the variables assuming no correlation between the measurement types.
\begin{table}[H]
	\caption{Variables for analytical Hessian and Jacobian matrix of \acrshort{ml} approach from (\ref{eq:likelihood}).}
	\label{tab:analytical_hessians}
	
\begin{tabular}{llllll}
    \toprule
    Fusion Type & Measurement Vector & Measurement Covariance \\
    \midrule
    \acrshort{aod}+\acrshort{aoa} & $\tilde{\vect{m}}=\left[\tilde{\eta}_\mathrm{\acrshort{tx}}, \tilde{\eta}_\mathrm{\acrshort{rx}}\right]^\intercal$ & $\matr{C}=\diag\left(\sigma^2_\mathrm{\eta}, \sigma^2_\mathrm{\eta}\right)$ \\
    \acrshort{aod}+R & $\tilde{\vect{m}}=\left[\tilde{\eta}_\mathrm{\acrshort{tx}}, \tilde{r}_\mathrm{b}\right]^\intercal$ & $\matr{C}=\diag\left(\sigma^2_\mathrm{\eta}, \sigma^2_\mathrm{r}\right)$ \\
    \acrshort{aoa}+R & $\tilde{\vect{m}}=\left[\tilde{\eta}_\mathrm{\acrshort{rx}}, \tilde{r}_\mathrm{b}\right]^\intercal$ & $\matr{C}=\diag\left(\sigma^2_\mathrm{\eta}, \sigma^2_\mathrm{r}\right)$ \\
    \acrshort{aod}+\acrshort{aoa}+R & $\tilde{\vect{m}}=\left[\tilde{\eta}_\mathrm{\acrshort{tx}}, \tilde{\eta}_\mathrm{\acrshort{rx}}, \tilde{r}_\mathrm{b}\right]^\intercal$ & $\matr{C}=\diag\left(\sigma^2_\mathrm{\eta}, \sigma^2_\mathrm{\eta}, \sigma^2_\mathrm{r}\right)$ \\
    \bottomrule
\end{tabular}

\end{table}
From \eqref{eq:range} -- \eqref{eq:naf}, the corresponding conversion function $\vect{f}\left(\vect{p}\right)$ for (\ref{eq:likelihood}) is then a combination of
\begin{align*}
    f_{r_\mathrm{b}}\left(\vect{p}\right) &= \sqrt{\left(p_\mathrm{x}+c\right)^2+p_\mathrm{y}^2}+        \sqrt{\left(p_\mathrm{x}-c\right)^2+p_\mathrm{y}^2} \, ,\\
    f_{\eta_\mathrm{\acrshort{tx}}}\left(\vect{p}\right) &= -\frac{d}{\lambda}\sin\left(\arctan\left(\frac{p_\mathrm{x} + c}{p_\mathrm{y}}\right)\right) \, , \quad p_\mathrm{y} > 0 \, , \\
    f_{\eta_\mathrm{\acrshort{rx}}}\left(\vect{p}\right) &= -\frac{d}{\lambda}\sin\left(\arctan\left(\frac{p_\mathrm{x}-c}{p_\mathrm{y}}\right)\right) \, , \quad p_\mathrm{y} > 0 \, . \\
\end{align*}

\end{appendices}

\bibliography{references}
\bibliographystyle{IEEEtran}

\end{document}